\documentclass{amsart}
\usepackage{graphicx} 
\usepackage{hyperref}
\usepackage{nameref,hyperref}
\usepackage{url}
\usepackage{subfig}
\usepackage{morefloats}
\usepackage{multirow}
\usepackage{palatino}
\usepackage{eucal}
\usepackage{todonotes}
\usepackage{setspace}
\usepackage{cite}
\usepackage{pbox}

\makeatletter
\renewcommand{\@biblabel}[1]{\quad#1.}
\makeatother
\pdfoutput=1

\newcommand{\imgt}{\textsf{IMGT}}
\newcommand{\partis}{\textsf{partis}}
\newcommand{\vsearch}{\textsf{vsearch}}
\newcommand{\ham}{\textsf{ham}}

\newcommand{\ighutil}{\textsf{ighutil}}
\newcommand{\changeo}{\textsf{Change-O}}
\newcommand{\mixcr}{\textsf{MiXCR}}

\newcommand{\PP}{\mathbb{P}}

\newcommand{\TP}{\operatorname{TP}}

\newcommand{\FP}{\operatorname{FP}}
\newcommand{\FN}{\operatorname{FN}}
\newcommand{\sensitivity}{\operatorname{Sensitivity}}
\newcommand{\precision}{\operatorname{Precision}}

\newcommand{\vdj}{VDJ}
\newcommand{\nbiggestclusters}{40}
\newcommand{\ccs}{candidate cluster size}
\newcommand{\supplementUrl}{\textsf{figshare} at \url{http://figshare.com/s/9b85e4ac54d011e5bd3e06ec4b8d1f61}}

\newcommand{\forarxiv}[1]{#1}
\newcommand{\notforarxiv}[1]{}
\newcommand{\beginsupplement}{%
        \setcounter{table}{0}
        \renewcommand{\thetable}{S\arabic{table}}%
        \setcounter{figure}{0}
        \renewcommand{\thefigure}{S\arabic{figure}}%
     }

\newcommand{\similarityMatrixExplain}{
  For these plots, we took the \nbiggestclusters\ largest clusters resulting from the given clustering and took their intersection with the \nbiggestclusters\ largest clusters generated by the simulation.
  Each non-white square indicates that there was a non-empty intersection between the two clusters; the square is shaded by the size of the clusters' intersection divided by their mean size.
  The position of the square shows the relative sizes of the two clusters.
}
\newcommand{\ccfFigExplain}{
  via per-read averages of precision (top left), sensitivity (top right), and their harmonic mean (bottom, called F1 score).
  Results are on simulated sequences which span the entire V, D, and J segments; the number of leaves (BCR sequences per clonal family) is distributed geometrically with the indicated mean value.
  Precision measures the extent to which inferred clusters contain truly clonal sequences, while sensitivity measures the extent to which the entirety of each sequence's clonal family appears in its inferred cluster.
}

\title{Likelihood-based inference of B cell clonal families}
\author[Ralph]{Duncan K. Ralph}
\author[Matsen]{Frederick A. Matsen IV}

\begin{document}

\begin{abstract}
The human immune system depends on a highly diverse collection of antibody-making B cells.
B cell receptor sequence diversity is generated by a random recombination process called ``rearrangement'' forming progenitor B cells, then a Darwinian process of lineage diversification and selection called ``affinity maturation.''
The resulting receptors can be sequenced in high throughput for research and diagnostics.
Such a collection of sequences contains a mixture of various lineages, each of which may be quite numerous, or may consist of only a single member.
As a step to understanding the process and result of this diversification, one may wish to reconstruct lineage membership, i.e. to cluster sampled sequences according to which came from the same rearrangement events.
We call this clustering problem ``clonal family inference.''
In this paper we describe and validate a likelihood-based framework for clonal family inference based on a multi-hidden Markov Model (multi-HMM) framework for B cell receptor sequences.
We describe an agglomerative algorithm to find a maximum likelihood clustering, two approximate algorithms with various trade-offs of speed versus accuracy, and a third, fast algorithm for finding specific lineages.
We show that under simulation these algorithms greatly improve upon existing clonal family inference methods, and that they also give significantly different clusters than previous methods when applied to two real data sets.
\end{abstract}


\maketitle
\notforarxiv{
\section*{Summary statement}
Antibodies must recognize a great diversity of antigens to protect us from infectious disease.
The binding properties of antibodies are determined by the sequences of their corresponding B cell receptors (BCRs).
These BCR sequences are created in ``draft'' form by VDJ recombination, which randomly selects and deletes from the ends of V, D, and J genes, then joins them together with additional random nucleotides.
If they pass initial screening and bind an antigen, these sequences then undergo an evolutionary process of reproduction, mutation, and selection, ``revising'' the BCR to improve binding to its cognate antigen.
It has recently become possible to determine the BCR sequences resulting from this process, which determine antibody binding, in high throughput.
Although these sequences implicitly contain a wealth of information about both antigen exposure and the process by which we learn to resist pathogens, this information can only be extracted using computer algorithms.
In this paper we describe a likelihood-based statistical method to determine, given a collection of BCR sequences, which of them are derived from the same recombination events.
It is based on a hidden Markov model (HMM) of VDJ rearrangement which is able to calculate likelihoods for many sequences at once.
}

\section*{Introduction}
B cells effect the antibody-mediated component of the adaptive immune system.
The antigen-binding properties of B cells are defined by their B cell receptor, or BCR.
BCRs bind a wide variety of antigens, and this flexibility arises from their developmental pathway.
B cells begin life as hematopoietic stem cells.
After a number of differentiation steps the cells perform somatic recombination, or rearrangement.
For the heavy chain locus, a V gene, D gene, and J gene are randomly selected, trimmed some random amount by an exonuclease, and then joined together with random nucleotides (forming so-called N-regions).
The light chain process is slightly simpler, in that only a V and J recombine, but proceeds via similar trimming and joining processes.
These processes form the third complementarity determining region (CDR3) in each of the heavy and the light chain, which are important determinants of antibody binding properties.
Then a series of checkpoints on the BCRs ensure that the resulting immunoglobulin is functional and not self-reactive through negative selection (reviewed in \cite{Melchers2015-co}).
This process results in naive B cells with fully functioning receptors.
When stimulated by binding to antigen in a germinal center, naive cells reproduce and mutate by via the process of somatic hypermutation, and then are selected on the basis of antigen binding and presentation to T follicular helper cells \cite{Victora2014-hm}.
This process is called \emph{affinity maturation}.
It is now possible to sequence B cell receptors in high throughput, which in principle describes not only the collections of antigens to which the immune system is ready to react, but also implicitly narrates how they came to be.

It is of great practical interest for researchers to be able to reconstruct events of this development process using BCR sequence data.
Such reconstruction would shed light on the process of B cell receptor maturation, a subject of continual study since the landmark work of Eisen and Siskind in 1964\cite{Eisen1964-ls,Cooper2015-pk}.
Furthermore, there are specific maturation pathways of great importance, such as the B cell lineages leading to broadly neutralizing antibodies to HIV \cite{Doria-Rose2014-vi,Wu2015-uw}.
Being able to reconstruct the structure and history of these lineages allows investigation of the binding properties of these intermediates, which could be helpful to design effective vaccination strategies to elicit high-affinity antibodies \cite{Mascola2013-mt}.
For example, recent studies have shown the promise of a sequential immunization program for eliciting these antibodies \cite{Dosenovic2015-ek}; lineage reconstruction will aid in identifying desirable intermediate BCRs.

\begin{figure}[!ht]
\forarxiv{\includegraphics[width=2.8in]{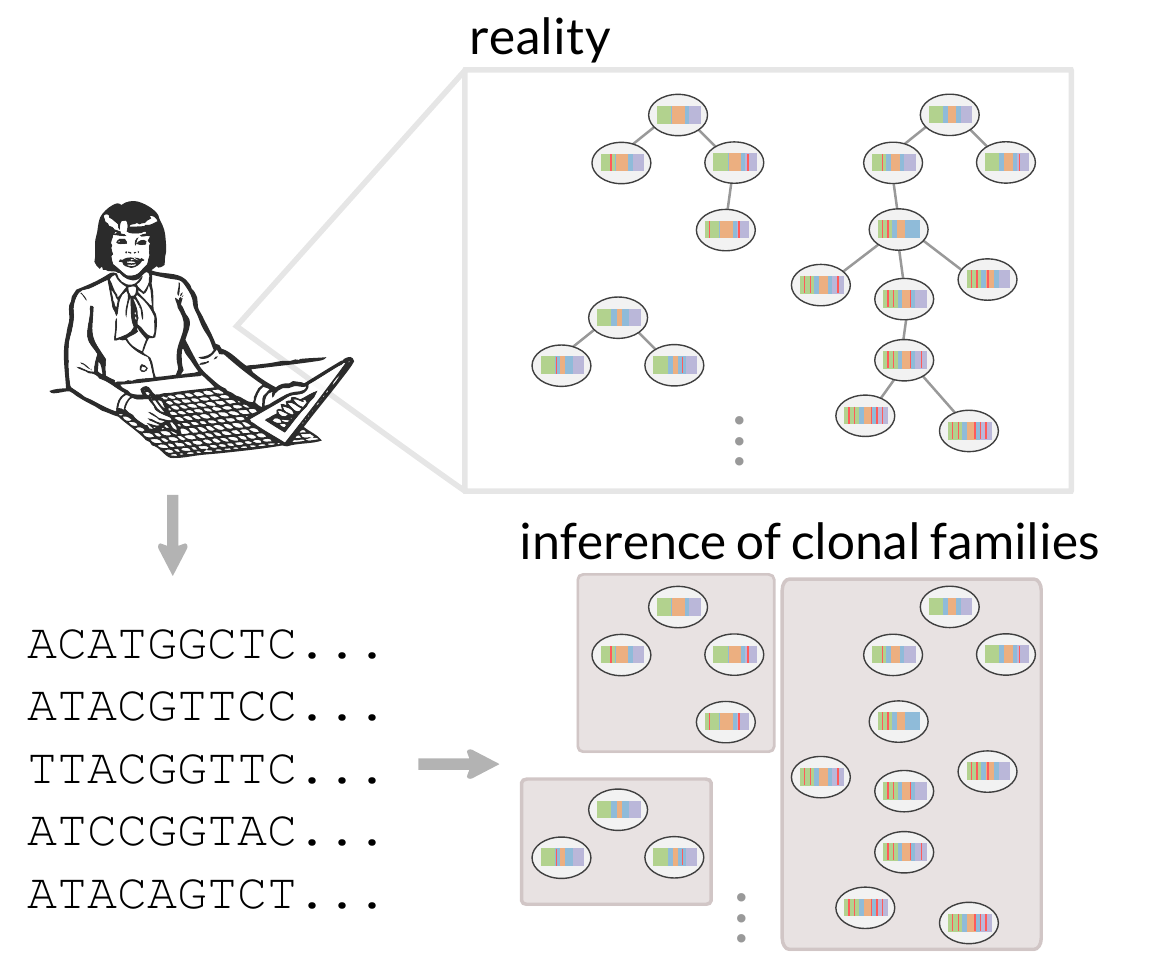}}
\caption{\
  {\bf The clonal family inference problem.}
  The B cell receptor generation process begins by VDJ recombination, which makes a naive B cell.
  When stimulated by antigen, those naive cells diversify through the mutation and selection processes of \emph{affinity maturation}, creating many lineages of B cells shown here as phylogenetic trees with the naive cells at the root of the tree.
  The ensemble of B cells descending from a single rearrangement event is called a \emph{clonal family}.
  In this paper we develop methods to reconstruct clonal families from B cell receptor sequences.
}
\label{FIGproblem}
\end{figure}

The \emph{clonal family inference problem} is an intermediate step to such lineage reconstruction.
Rather than trying to reconstruct the full lineage history of the set of sequences, the goal is only to reconstruct which sequences came from the same rearrangement event.
Full lineage reconstruction would also require building phylogenetic trees for each of the clonal families.
However, these clonal families can be an object of interest themselves \cite{Bashford-Rogers2013-xv}.

The motivation behind our approach to the clonal family inference problem, like many before us, is to use the special structure of BCR sequences (which for simplicity we describe for the heavy chain; the same concepts and approaches can be applied to the light chain).
This structure follows from VDJ recombination and affinity maturation: for example, by definition the identity of the germline genes cannot change through affinity maturation.
Thus, if the per-read germline gene identity could be inferred without error, then any pair of sequences from a clonal family must have the same inferred germline gene identity.
If one also assumes that sequences evolve only through point mutation, then sequences must have identical-length CDR3s if they are to be in the same clonal family.

Most current methods for B cell clonal family inference make these assumptions, and proceed by first stratifying sequences by inferred V and J germline genes and CDR3 length, then only consider pairs of sequences within a stratum as potential members of the same clonal family.
If one assumes further that any clonal families with pairs of highly diverged sequences also contain intermediates between those sequences, one might assume that there is a path between any pair of sequences such that neighboring sequences in a path are similar.
This suggests a strategy in which pairs of sequences that are similar at some level (such as 90\% similar in terms of nucleotides) in the CDR3 are considered to be in the same clonal family, and where membership is transitive, which corresponds to an application of single-linkage clustering.

Instead of designing such an algorithm that works only when a set of rigid, predefined assumptions are satisfied, an alternative is to formalize a model of B cell affinity maturation into a generative probabilistic process with a corresponding likelihood function.
Once this likelihood function is defined, one can infer clonal families by finding the clustering that maximizes the likelihood of generating the observed sequences.

Likelihood methods in the form of a hidden Markov models (HMM) have been applied to B cell receptor sequences for a decade \cite{Volpe2005-uk,Gaeta2007-mz,Munshaw2010-mj,Elhanati2016-yq}.
This previous work has been to use HMMs to analyze individual sequences.
For likelihood-based clustering we are only aware of the work of Laserson \cite{Laserson2012-pi,Laserson2014-yh}, who uses Markov chain Monte Carlo to infer clusters via a Dirichlet mixture model (reviewed in \cite{Neal2000-hi}).
Unfortunately the Laserson algorithm is only described in a PhD thesis and does not appear to be publicly available.
In related work, Kepler \cite{Kepler2013-sy,Kepler2014-jy} uses a likelihood-based phylogenetics framework to perform joint reconstruction of annotated ancestor sequence and a phylogenetic tree.

In this paper we present a method for inferring clonal families in an HMM-based framework that comfortably scales to tens of thousands of sequences via parallel algorithms, with approximations that scale to hundreds of thousands of sequences.
For situations in which specific lineages are of interest, users can specify ``seed'' sequences and find the clonal family containing that seed in repertoires with one million sequences.
Our clustering algorithm is based on a ``multi-HMM'' framework for BCR sequences that we have previously applied to the \emph{annotation problem}: to infer the origin of each nucleotide in a BCR (or TCR) sequence from the VDJ rearrangement process \cite{Ralph2016-kr}.
We use this framework to define a likelihood ratio comparing two models which differ by the collapse of two clonal families into one, and use it for agglomerative clustering.
Because this likelihood ratio comes from an application of the forward algorithm for HMMs, it integrates out all possible VDJ annotations.
We find that it outperforms previous algorithms on simulated data, and that it makes a significant difference when applied to real data.

\section*{Results}

\subsection*{Likelihood framework}
In order to calculate a set of probabilities suitable for use in the clonal family inference problem, we begin with the HMM framework introduced in~\cite{Ralph2016-kr}.
In that paper we focused on inferring parameters of an HMM and using it to obtain BCR annotated ancestor sequences, which was primarily based on the most likely path through each HMM, i.e.\ the Viterbi path.
We also described Viterbi annotation with a multi-HMM, i.e.\ annotation using a collection of sequences that were assumed to form a clonal family.

In this application, we will use the forward algorithm for HMMs~\cite{Durbin1998-uq} to obtain the corresponding marginal probability, which is the sum of sequence generation probabilities over all possible paths through the HMM.
This is a more appropriate tool for the clonal family inference problem because here we are interested in integrating over annotated ancestor sequences (that is, paths through the HMM) to decide whether sequences are related.
By using a multi-HMM, we can use this total probability to calculate a likelihood ratio that two clusters derive from the same, or from different, rearrangement events.
We perform agglomerative clustering using this likelihood ratio to group sequences for which the probability of a common ancestry is higher than that of separate ancestry (details in the Methods).
This approach allows us to calculate the total probability of the partition (i.e.\ clustering) at each stage in the clustering process, which provides both an objective measure of partition quality, and easy access to not only the most likely partition but also to a range of likely partitions of varying degrees of refinement.
As in our previous work, the parameters of the HMM can be inferred ``on the fly'' given a sufficiently large data set or be inferred on some other data set.
Briefly, we do a cycle of Viterbi training, which is started with an application of Smith-Waterman alignment, in which the best annotation for each sequence with a current parameter set is used to infer parameters for the next cycle.
As described in detail elsewhere \cite{Ralph2016-kr}, data is aggregated if there are insufficient observations for a given allele for training.

\subsection*{Approximate Methods}
In addition to this principled method for full-repertoire reconstruction, we have implemented two more approximate versions which trade some accuracy for substantial increases in speed.
In the first, which we call \emph{point \partis}, we forgo integration over all possible annotated ancestor sequences and instead find the most likely naive sequence point estimate for each cluster.
Clusters are then compared based on the Hamming fraction (Hamming distance divided by sequence length) between their respective naive sequences, and are merged if the distance is smaller than some threshold.
This threshold is set dynamically based on the observed mutation rate in the sample at hand.

In order to achieve further improvements in speed, we can also avoid both complete all-versus-all comparison of the sequences at each step, and calculation of the joint naive sequence for each merged cluster.
For this we find the most likely naive sequence for each individual sequence, and then pass the results, together with a dynamically-set clustering threshold, to the clustering functionality of the \vsearch\ program \cite{vsearch}.
We call this \emph{\vsearch\ \partis}.

\subsection*{Reconstruction of selected lineages}
We have also included a method which, using the full likelihood, reconstructs the clonal family containing a given ``seed'' sequence.
Because clonal families are generally significantly smaller than the total repertoire, this option is much faster than the full-repertoire reconstruction methods.
We see this option as being useful when specific sequences are identified as interesting through a binding assay or because they are shared between repertoire samples.
This is labeled \emph{full \partis\ (seed)}.

\subsection*{Implementation}
This clustering has been implemented as part of continued development of \partis\ (\url{http://github.com/psathyrella/partis}).
As before, the license is GPL v3, and we have made use of continuous integration and containerization via Docker for ease of use and reproducibility \cite{Boettiger2014-mm}.
A Docker image with \partis\ installed is available at \url{https://registry.hub.docker.com/u/psathyrella/partis/}.

\subsection*{Results on simulation}

\begin{figure}[!ht]
  \forarxiv{\includegraphics[width=5.5in]{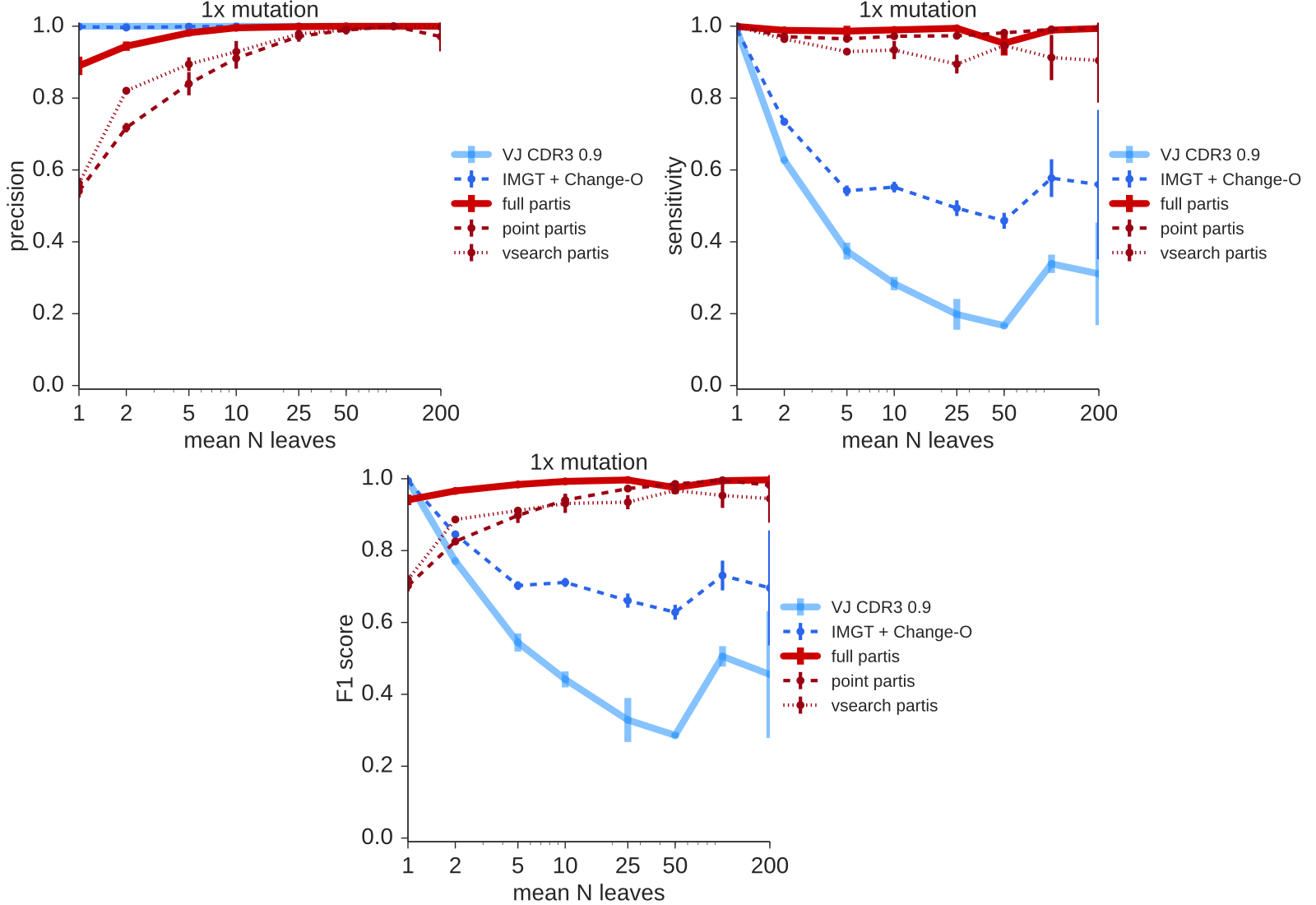}}
\caption{\
  {\bf Similarity between inferred and true partitions for the various clustering methods at typical ($1 \times$) mutation levels}
  \ccfFigExplain
}\label{FIGccfMut1NoSynthetic}
\end{figure}

\begin{figure}[!ht]
  \forarxiv{\includegraphics[width=5.5in]{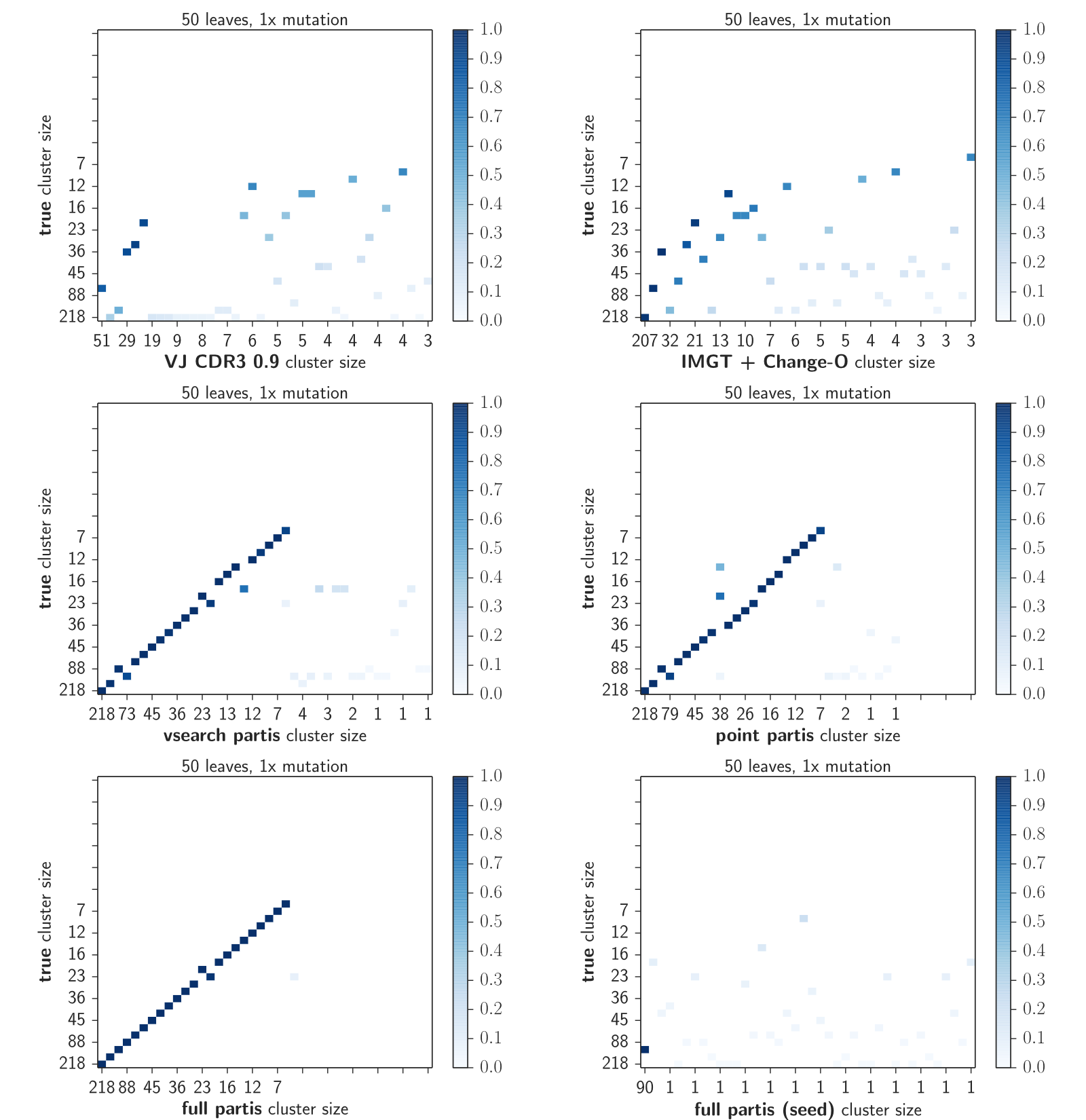}}
  \caption{\
  {\bf Fraction of sequences per cluster in common with the true partition on simulation for each method at typical ($1 \times$) levels of mutation.}
  \similarityMatrixExplain
  Results are shown for the simulation sample in which the size of each clonal family is drawn from a geometric distribution with mean 50 (other values are shown in Figures~\ref{FIGsimilarityMatricesSimu2LeavesMut1} and~\ref{FIGsimilarityMatricesSimu200LeavesMut1}).
}
\label{FIGsimilarityMatricesSimu50LeavesMut1}
\end{figure}

\begin{figure}[!ht]
  \forarxiv{\includegraphics[width=5.5in]{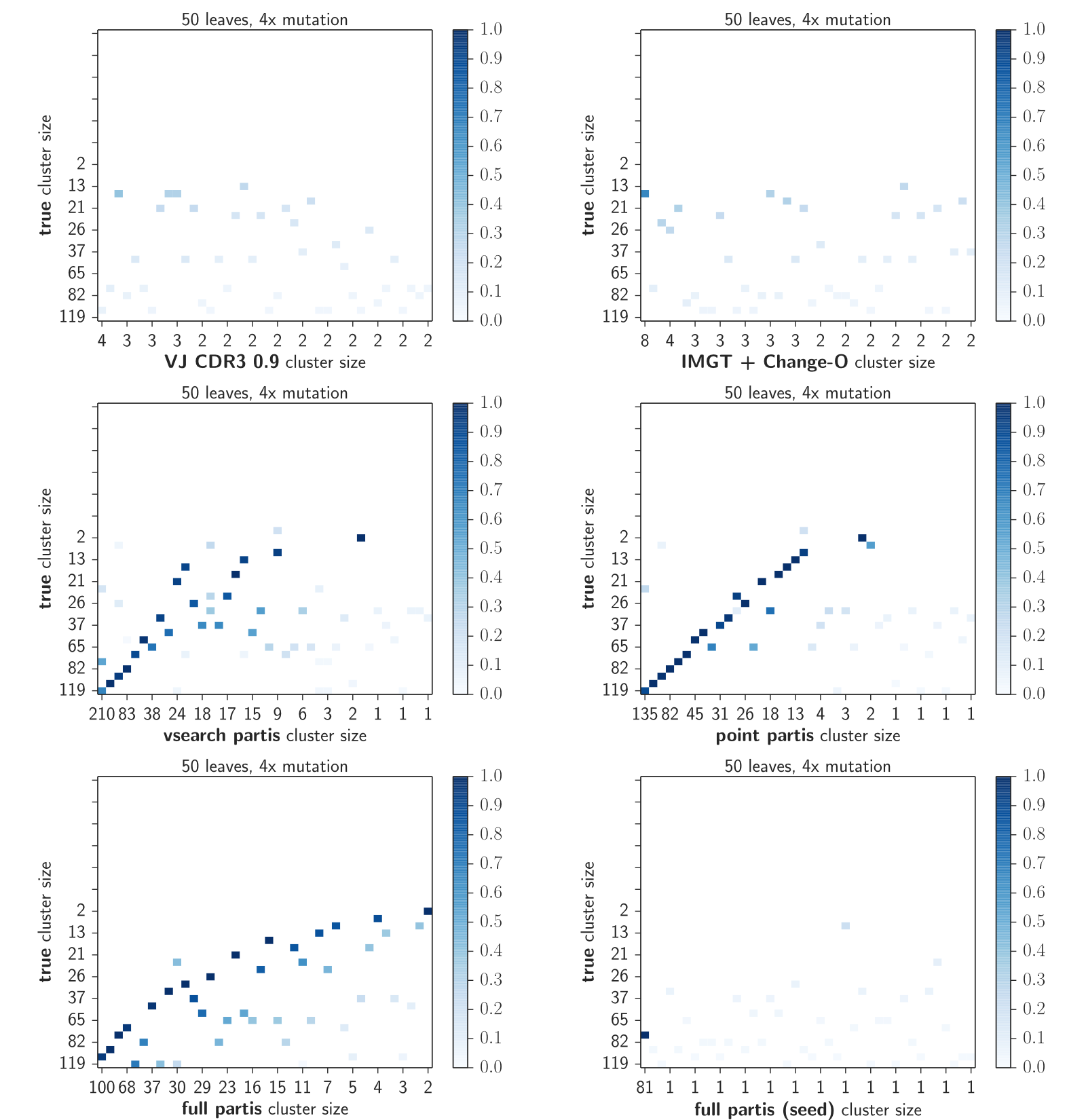}}
\caption{\
  {\bf Fraction of sequences per cluster in common with the true partition on simulation for each method with high ($4 \times$) mutation.}
  Plot layout as in Figure~\ref{FIGsimilarityMatricesSimu50LeavesMut1}.
  Results are shown for the simulation sample in which the size of each clonal family is drawn from a geometric distribution with mean 50 (other values are shown in Figures~\ref{FIGsimilarityMatricesSimu2LeavesMut4}, and~\ref{FIGsimilarityMatricesSimu200LeavesMut4}).
}
\label{FIGsimilarityMatricesSimu50LeavesMut4}
\end{figure}

\begin{figure}[!ht]
  \forarxiv{\includegraphics[width=5.5in]{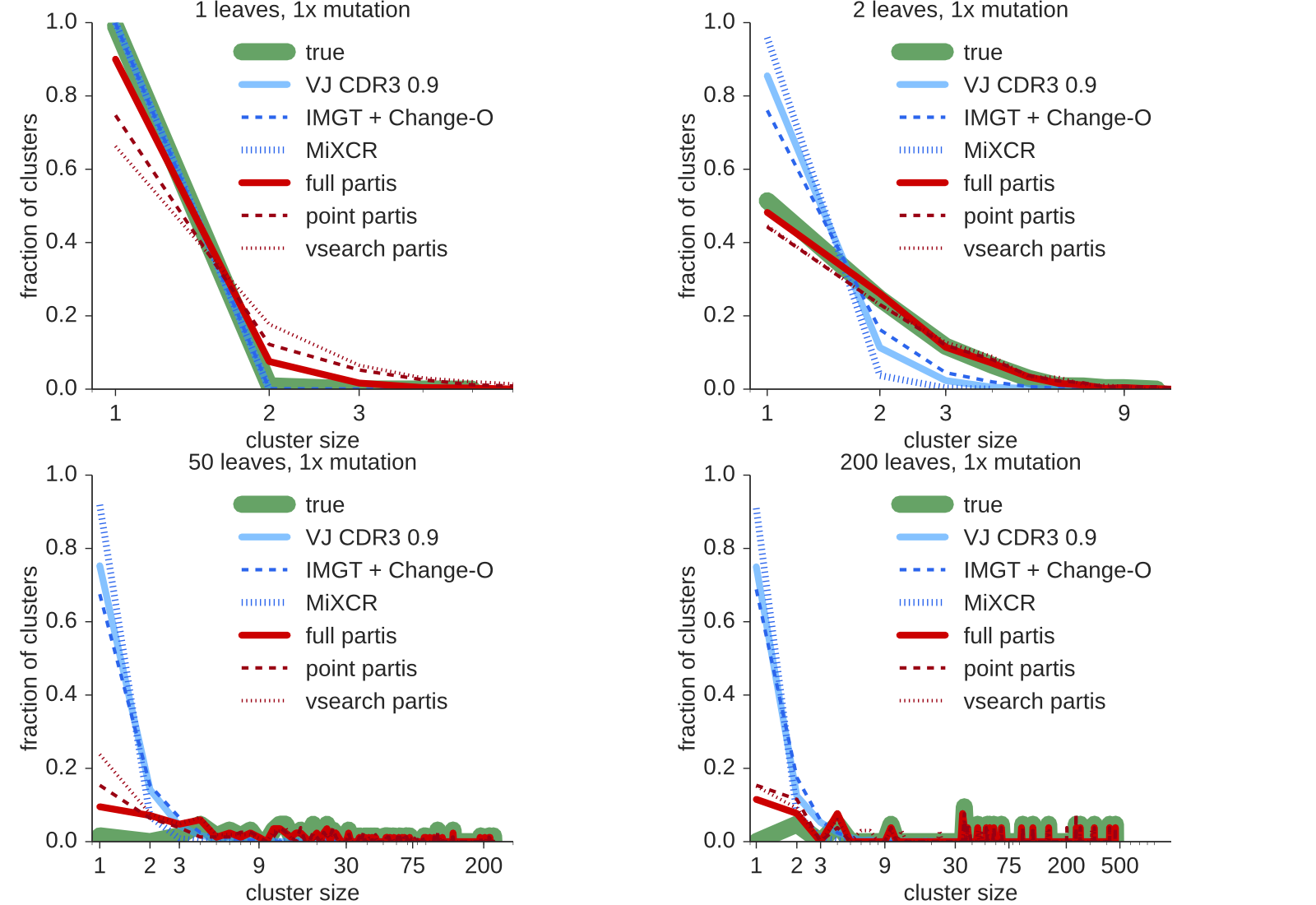}}
\caption{\
  {\bf True and inferred cluster size distributions at normal ($1 \times$) mutation levels for each of the methods} for geometrically distributed simulated cluster sizes with various means.
  Results are the mean of three simulated samples with 1000 sequences each.
}
\label{FIGclusterSizeMut1}
\end{figure}

\begin{figure}[!ht]
  \forarxiv{\includegraphics[width=5.5in]{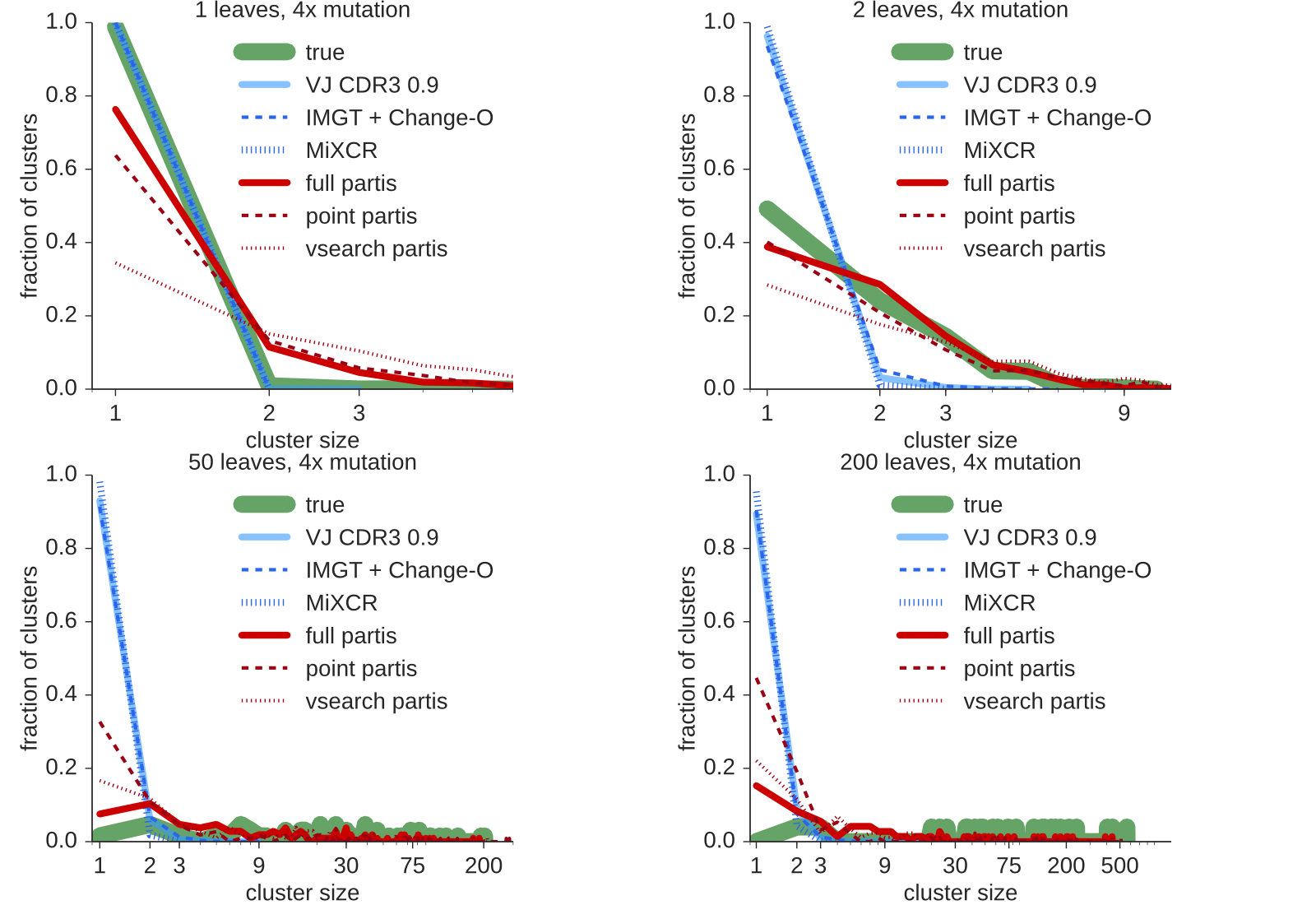}}
\caption{\
  {\bf True and inferred cluster size distributions at high mutation levels ($\times 4$) for each of the methods} for geometrically distributed simulated cluster sizes with various means.
  Results are the mean of three simulated samples with 1000 sequences each.
}
\label{FIGclusterSizeMut4}
\end{figure}

\begin{figure}[!ht]
  \forarxiv{\includegraphics[width=5.5in]{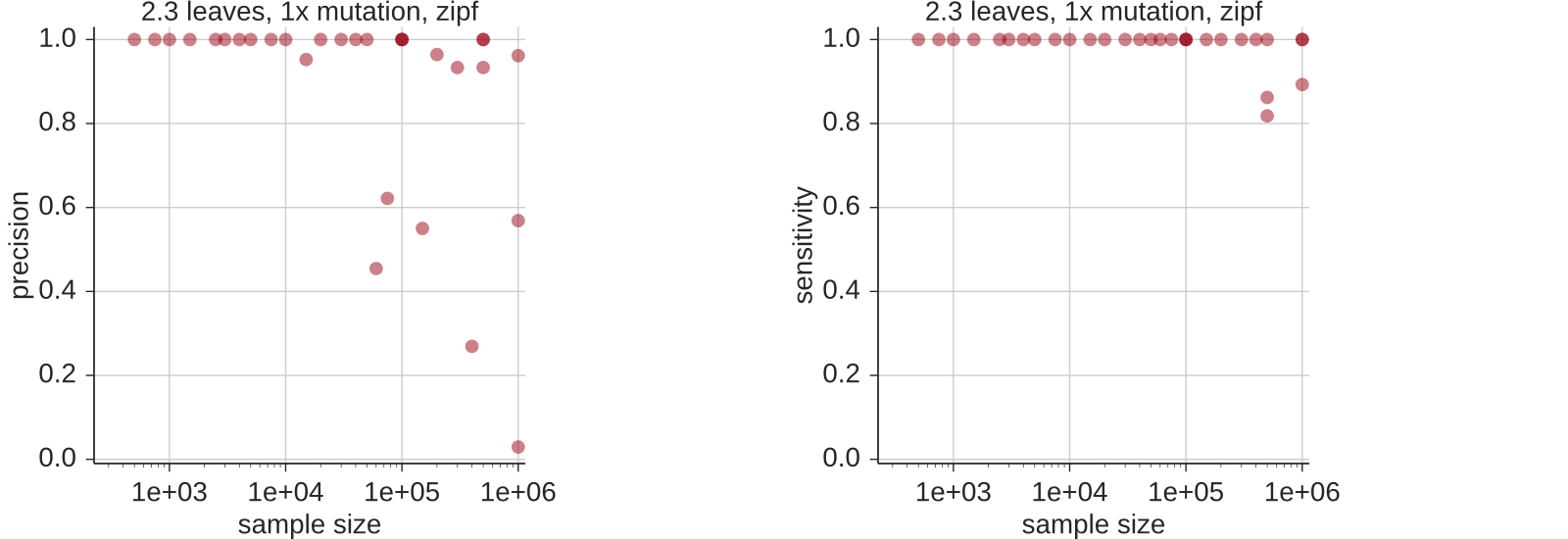}}
\caption{\
  {\bf Similarity between inferred and true partitions for seed \partis} via per-read averages of precision (left) and sensitivity (right) with increasing sample size.
  There is one point at each indicated $x$ value except for one hundred thousand, five hundred thousand, and one million, which have three points each.
  Results are shown for a sample with cluster sizes distributed as a Zipf (power law) distribution with exponent 2.3, given a randomly selected seed sequence from a randomly selected large cluster.}
\label{FIGseedCCFS}
\end{figure}

In the absence of real data sets with many sequences for which the true annotations and lineage structures are known, we compare these new clustering methods against previous methods using simulated sequences generated as described in~\cite{Ralph2016-kr}.
These simulations were done for the heavy chain locus only.
We performed comparison both on samples, which we call $1 \times$, which mimic mutation frequencies in data (overall mean frequency of about 10\%) and on samples, which we call $4 \times$, with quadrupled branch lengths (overall mean frequency of about 25\%) to explore results in a more challenging regime.
Per-sequence mutation frequencies are distributed according to the empirical distribution (see~\cite{Ralph2016-kr}).
We compare the three \partis\ methods to three methods from the literature.
The first, labeled ``VJ CDR3 0.9'', is representative of annotation- and distance-based methods which have been used in a number of papers~\cite{Jiang2011-ur,Vollmers2013-vh,Kepler2014-jy,Stern2014-ph,Yaari2015-ss}.
It begins by annotating each individual sequence, and proceeds to group sequences which share the same V and J gene and the same CDR3 length, and have CDR3 sequence similarity above some threshold, which is commonly 0.9~\cite{Vollmers2013-vh}.
For this comparison we use \partis\ annotation; for a comparison of annotation methods themselves see~\cite{Ralph2016-kr}.
We also compare against \changeo's clustering functionality \cite{Gupta2015-sd} fed with annotations from \imgt, with \imgt\ failures (when it does not return an annotation) classified as singletons.
We perform a partial comparison against \mixcr~\cite{Bolotin2015-zb}.
Since this method does not currently report which sequences go into which clusters, and instead only reports cluster summary statistics, we cannot perform a detailed evaluation.
The authors of \mixcr\ note in personal communication, however, that they plan to report this information in future versions.

We use per-read averages of precision and sensitivity to quantify clustering accuracy.
In this context, the precision for a given read is the fraction of sequences in its inferred cluster which are actually in its clonal family, while sensitivity for a given read is the fraction of sequences in its true clonal family that appear in its inferred cluster (details in Methods).
We find that \partis\ is much more sensitive than previous methods, at the cost of some loss of precision (Fig~\ref{FIGccfMut1NoSynthetic}).
The point \partis\ approximate implementation is less specific than the full implementation, while the even faster \vsearch\ approximation loses some precision and some sensitivity.

We investigate these differences in more detail for the first simulation replicate via an \emph{intersection matrix} with entries equal to the size of the intersection between each of the \nbiggestclusters\ largest clusters returned by pairs of algorithms (Figures~\ref{FIGsimilarityMatricesSimu50LeavesMut1} and~\ref{FIGsimilarityMatricesSimu50LeavesMut4}, and Figures~\ref{FIGsimilarityMatricesSimu2LeavesMut1},~\ref{FIGsimilarityMatricesSimu2LeavesMut4},~\ref{FIGsimilarityMatricesSimu200LeavesMut1}, and~\ref{FIGsimilarityMatricesSimu200LeavesMut4}).
Full \partis\ infers clonal families correctly the majority of the time at typical mutation levels, and in this experiment it incorrectly split a cluster of true size around 45.
These results degraded somewhat with the point \partis\ approximation, and somewhat more with the \vsearch\ approximation.
The VJ CDR3 0.9 method consistently under-clustered for the largest cluster sizes.
The seeded full \partis\ method correctly reconstructed the lineage of interest starting from a randomly sampled sequence, while ignoring all others.

In order to understand performance on the many smaller clusters and to get a simpler overall picture, we also compared cluster size distributions for the various methods with the simulated distribution (Figs~\ref{FIGclusterSizeMut1},~\ref{FIGclusterSizeMut4},~\ref{FIGclusterSizeBox}, and~\ref{FIGclusterSizeMimic}).
Here we can see that \partis\ is able to accurately infer the true cluster size in a variety of regimes, whereas other methods tend to under-merge clusters of all sizes.

In order to further understand the source of these differences, we also compare results against two methods of generating incorrect partitions starting from the true partition, which we call \emph{synthetic} partitions (Figs~\ref{FIGccfMut1} and ~\ref{FIGccfMut4}).
The first, called \emph{synthetic 60\% singleton} is generated from the true partition by splitting 60\% of the sequences into singleton clusters.
The second, called \emph{synthetic neighbor 0.03}, merges together true clonal families which have true naive sequences closer than 0.03 in Hamming distance divided by sequence length.
We find that the performance of synthetic 60\% singleton tracks that of the VJ CDR3 method, while the performance of synthetic neighbor 0.03 tracks that of \partis.

Finally, to investigate the performance of the seeded full \partis\ method, we calculate the precision and sensitivity of this method on a number of widely varying sample sizes (Fig~\ref{FIGseedCCFS}).
For these simulations we used a Zipf (power-law) distribution of cluster sizes with exponent 2.3, and randomly selected one seed sequence from a randomly selected large cluster.
We find that seeded \partis\ frequently obtains very high sensitivity, although precision decreases as sample size increases.
This precision decrease is from incorrect merges of clusters.
We have manually checked these incorrect merges, and found that the true (i.e.\ simulated) naive sequences of clusters which are incorrectly merged with the seeded cluster typically differ by one to six bases.
Because these differences occur either within the bounds of the true eroded D segment, or within the true non-templated insertions, it is difficult to distinguish them from somatic hypermutation.
This echoes the observation that \partis\ precision is driven by the presence or absence of clusters which stem from different rearrangements, but which are very similar in naive sequence (compare \partis\ and synthetic neighbor 0.03 in Fig~\ref{FIGmetricsVsSampleSize}).

\subsection*{Insertion and deletion mutations}
\begin{figure}[!ht]
  \forarxiv{\includegraphics[width=5.5in]{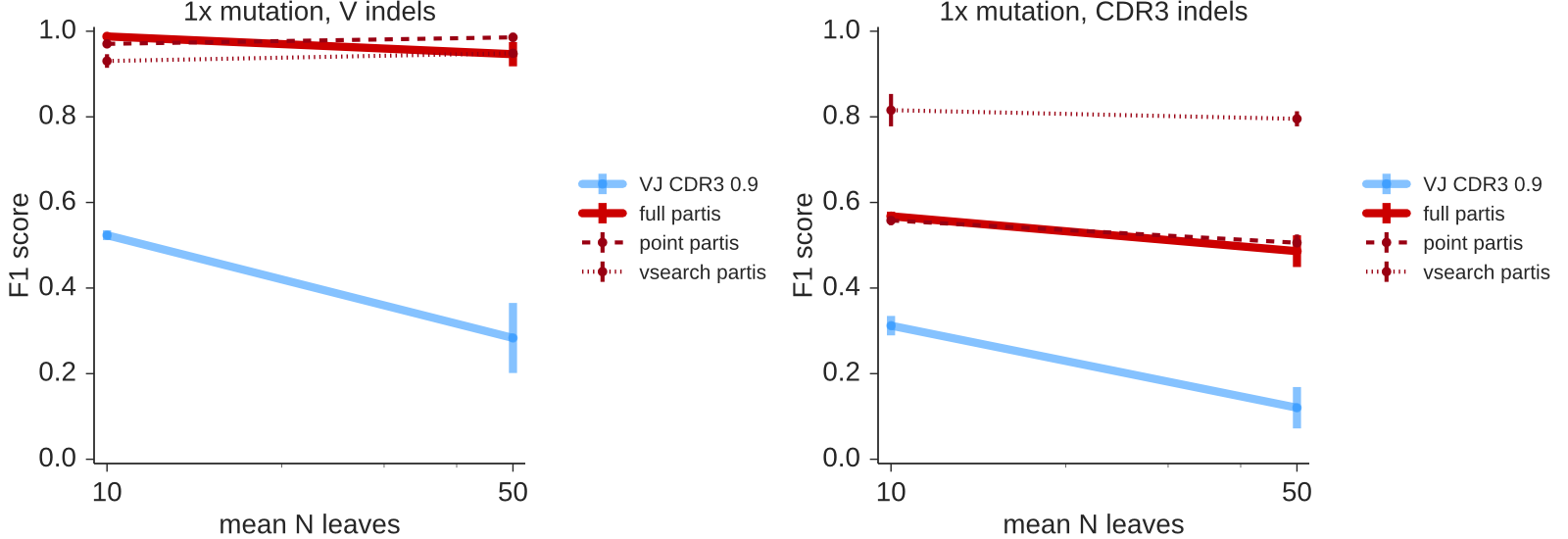}}
\caption{\
  {\bf Overall clustering quality, parameterized by the harmonic mean of precision and sensitivity (the F1 score), in the presence of indel mutations.}
  For these simulations, half of the simulated sequences have a single indel, whose position is distributed evenly either in the V segment (left, specifically between position 10 and the conserved cysteine) or in the CDR3 (right).
  Indels substantially decrease performance only when they occur within the CDR3.
  Results are the mean of 3 samples of 1000 sequences each.
}
\label{FIGmetricsIndels}
\end{figure}

In order to handle insertion-deletion (indel) mutations which occur during somatic hypermutation, we have implemented a heuristic method in the preliminary Smith-Waterman alignment step in \partis.
In short, this works by ``reversing'' inferred indel mutations in germline-encoded regions and proceeding with the clustering algorithm.
We find that \partis\ performance is typically unaffected when indels occur in non-CDR3 germline-encoded regions, although performance suffers when indels occur in the CDR3 (Fig~\ref{FIGmetricsIndels}).
This is because indel mutations in the CDR3 are quite difficult to distinguish from insertions and deletions stemming from the VDJ rearrangement process using indel-handling schemes (such as ours) that only take one sequence at a time into account.

\subsection*{Application to data}
\begin{figure}[!ht]
  \forarxiv{\includegraphics[width=5.5in]{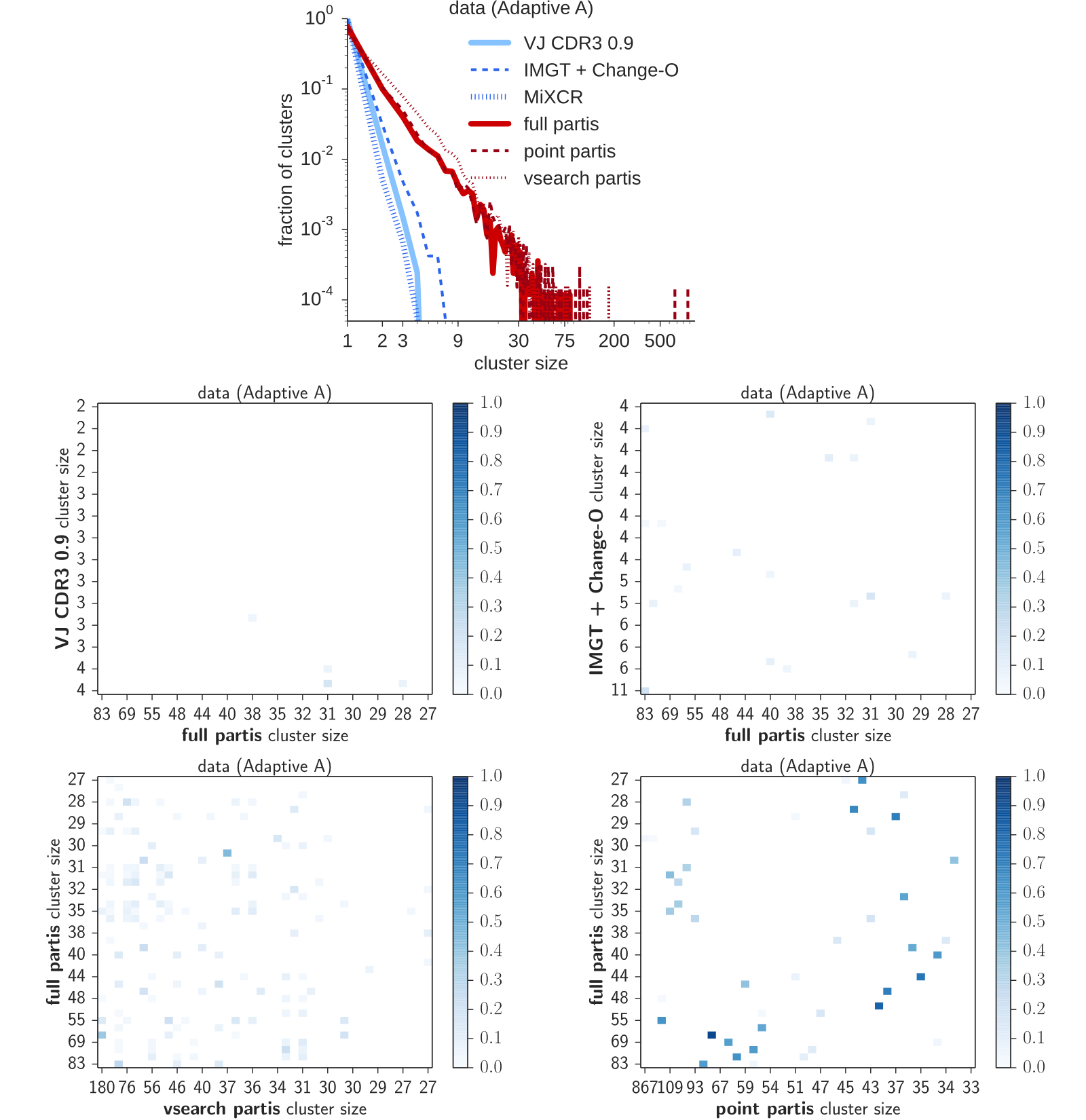}}
\caption{\
  {\bf Results of the various methods on data from subject A in the Adaptive data set.}
  We show cluster size distributions (top) and intersection matrices, which show the fraction of sequences per cluster in common between the various methods.
  Results are on a randomly-chosen subsample of 20,000 sequences.
}
\label{FIGadaptive}
\end{figure}

\begin{figure}[!ht]
  \forarxiv{\includegraphics[width=5.5in]{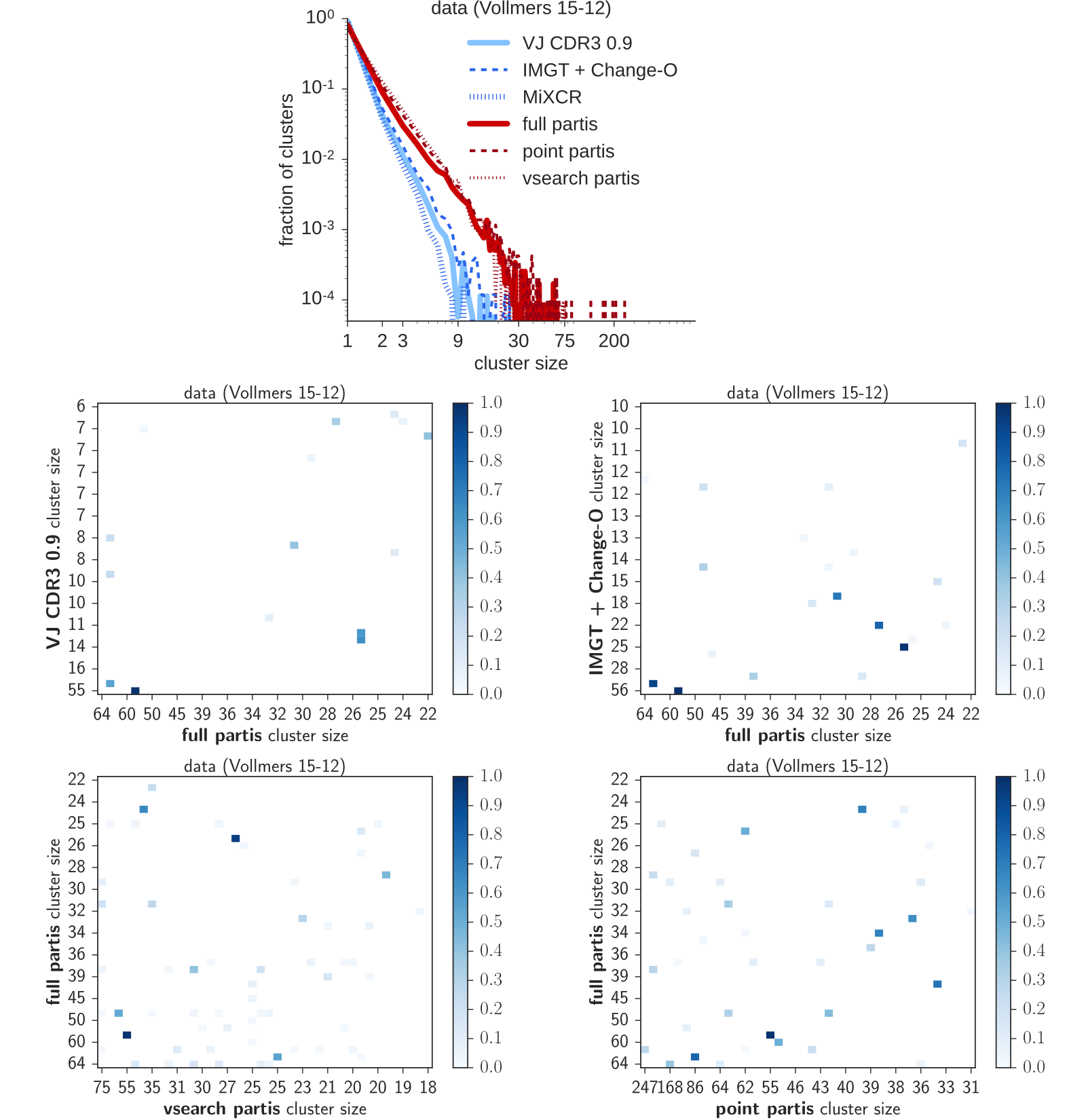}}
\caption{\
  {\bf Results of the various methods on data from subject 15-12 in the Vollmers data set.}
  We show cluster size distributions (top) and intersection matrices, which show the fraction of sequences per cluster in common between the various methods.
  Results are on a randomly-chosen subsample of 20,000 sequences.
}
\label{FIGstanford}
\end{figure}

In order to understand the difference this method makes on real data, we applied \partis\ and the other algorithms to subjects in the Adaptive data set from~\cite{billion} used in previous publications \cite{Ralph2016-kr,McCoy2015-qi,Elhanati2015-ld}, as well as the data set from \cite{Vollmers2013-vh}, which we will call the ``Vollmers'' data set.
These data sets were Illumina sequenced via amplicons covering the heavy chain CDR3, and thus do not have complete V or J sequences.
Especially in the case of the V region for the Vollmers data, it is not possible to confidently identify the germline V gene for each of the BCR sequences.
Thus, these data sets make for an interesting comparison between methods (such as VJ CDR3) which require single germline gene identifications, to our method, which integrates over such identifications.
Results are shown for Adaptive subject A (Fig~\ref{FIGadaptive}), and for a subject from the Vollmers data set (Fig~\ref{FIGstanford}).
The rest may be found on \supplementUrl.
Note that the identifiers shown for the Vollmers data are an obfuscated version of the original identifiers in the data; contact the authors for more details.
These results are not presented to make any strong statement about the true cluster size distribution, the correctness of which cannot be be independently evaluated, but rather to show that the \partis\ results are different from those of other methods on real data, as seen under simulation.

When we applied the various methods to a randomly chosen set of 20,000 sequences from two different sets, we found that the various methods agree that both samples are dominated by singletons, but there is substantial discord at the high end of the distribution, especially in Adaptive subject A (Figs~\ref{FIGadaptive} and~\ref{FIGstanford}).
These differences in composition are examined in more detail using cluster intersection matrices.
The cluster size distribution inferred by \partis\ approximately follows a power-law, with exponent about 2.3.

Adaptive subject A (Fig~\ref{FIGadaptive}) has mutation levels two and a half times higher than Vollmers subject 15-12 (Fig~\ref{FIGstanford}), making inference more challenging for A.
Both of these data sets consist of shorter sequences than the simulated sequences, which contain the entire V and J regions.
Reads in the Adaptive samples are 130 base pairs (losing about two thirds of the V and one half of the J), while those in the Vollmers data set vary in length, but typically span all of the J but only 20 to 30 bases in the V.

\subsection*{Time required}

\begin{figure}[!ht]
  \forarxiv{\includegraphics[width=3.5in]{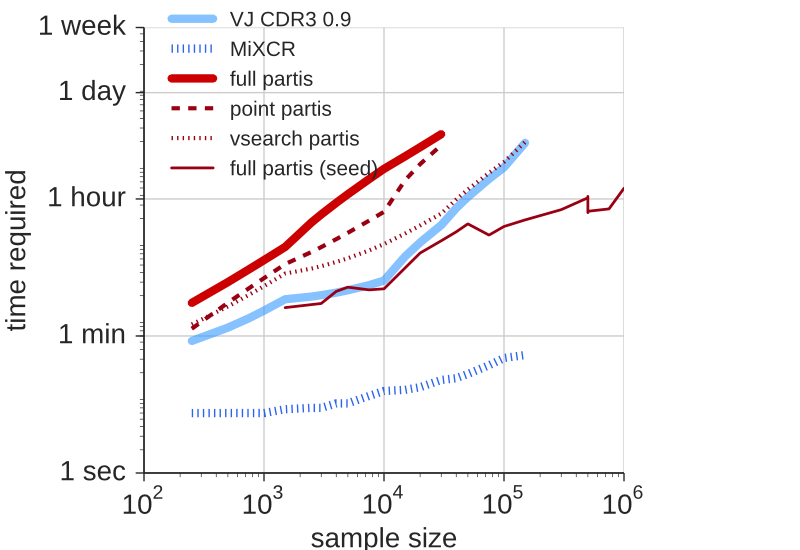}}
\caption{\
  {\bf Run time for the various methods}.
  Results are from running on a cluster with about 25 8-core machines.
  The time required for \changeo\ is difficult to measure, as the sequences are first annotated by manual submission to the IMGT website, which takes from 1-6 days to return results.
  The actual clustering time for \changeo\ once these annotations are obtained is very small, on par with the \mixcr\ results shown in this plot.
  Time required also varies by an order of magnitude depending on the structure of the sample (cluster size and mutation level).
}
\label{FIGtimeRequired}
\end{figure}

Likelihood-based clustering using \partis\ is computationally demanding, though within a range applicable to real questions given appropriate computing power (Fig~\ref{FIGtimeRequired}).
On a computing cluster with about 25 8-core machines, full and point \partis\ can cluster ten thousand sequences in 4 to 7 hours, while \vsearch\ \partis can cluster one hundred thousand sequences in 4 hours.
Our implementation of ``VJ CDR3 0.9'' used \partis\ annotation, but this approach could be made much faster by using a fast method for annotation \cite{Bolotin2015-zb,Kuchenbecker2015-hw}.
Time required can also vary by an order of magnitude depending on the structure of the sample (cluster size and mutation level).


\section*{Discussion}
We have developed an algorithm to infer clonal families using a likelihood-based framework.
Although the framework does take annotation information into account by using a VDJ-based HMM, the algorithm is distinguished from other clustering methods in that it does not fix a single annotation first and then use that annotation for downstream steps.
Instead we find that by integrating over annotated ancestor sequences using an HMM, we are able to obtain better clonal family inference than with the current common practice of rigidly inferring VJ annotation and then clustering on HCDR3 identity for heavy chain sequences.
Our simulations show that existing algorithms frequently do not sufficiently cluster sequences which sit in the same clonal family.
Our application to real data shows that the \partis\ algorithms using our default clustering thresholds return more large clusters on two real data sets, indicating that this difference in clustering is not simply an artifact of our simulation setup.

The performance differences between our various approximate algorithms indicates the sources of the \partis' improved performance.
The reasonably good performance of the point \partis\ variant shows the importance of clustering on inferred naive sequences rather than observed sequences and inferring these naive sequences with an accurate probabilistic method.
Furthermore, the difference between point and full \partis\ is some measure of the importance of integrating out uncertainty in annotated ancestor sequences.

We find that \partis' main weakness is in separating out clusters with highly similar naive sequences.
Indeed, its performance tracks a simulated method that merges clonal families with true (i.e.\ simulated) naive sequences that are closer than 3\% in nucleotides, in simulations with about 10\% divergence from the naive sequence.
Although the VDJ rearrangement process generates a very diverse repertoire, biases in gene family use and other rearrangement parameters mean that pairs of highly similar naive sequences are frequently generated.
This may indicate an inherent limitation in clonal family inference methods that only use data from heavy chain.

Our method builds on previous work for doing likelihood-based analysis of BCR sequences.
In particular, we are indebted to Tom Kepler for initiating the use of HMMs in BCR sequence analysis \cite{Volpe2005-uk} and for developing likelihood-based methods to infer unmutated common ancestor sequences while integrating over rearrangement uncertainty \cite{Bonsignori2011-ky,Kepler2013-sy,Liao2013-cr,Kepler2014-jy,Gao2014-ls}.

We did not compare to several related methods that have been described in the literature.
\textsf{IMSEQ}~\cite{Kuchenbecker2015-hw} is a recent method which is reported to be quite fast; however the current version appears mainly aimed toward T cell receptors, as it does not handle somatic hypermutation.
As it clusters based on V and J genes and 100\% CDR3 similarity, it is equivalent to the annotation-based method described above, except with a threshold inappropriate to B cells.
\textsf{Cloanalyst} performs joint reconstruction of annotated ancestor sequence and a phylogenetic tree given a collection of sequences assumed to form a clonal family \cite{Kepler2013-sy}.
\textsf{Immunitree} apparently uses a Dirichlet process mixture model for clustering, however, the algorithm is only fully described in a PhD thesis~\cite{Laserson2012-pi}, and does not appear to be publicly available (note that \url{https://github.com/laserson/vdj} performs straightforward single-linkage clustering and is in fact written by a sibling of the \textsf{Immunitree} author).
\textsf{IgSCUEAL}~\cite{Frost2015-rp} is a recent method that performs annotation and clustering using a phylogenetic approach.
Its clustering algorithm, however, is not part of the public distribution and is apparently undergoing revision.

There are several opportunities to improve \partis.
First, our current approach requires likelihood ratios to exceed a value based on cluster size; these cluster sizes are based on observing distributions of likelihood ratios under simulation.
A more principled approach would be preferable.
Second, our approach to insertion-deletion mutations in affinity maturation only uses one sequence at a time.
Thus it has an inherent difficulty differentiating between mutations in the course of affinity maturation versus insertion-deletion events that are part of VDJ rearrangement.
Third, our current code is only for the heavy chain alone or the light chain chain alone.
Extending the work to paired heavy and light chain BCR data is conceptually straightforward, although will require additional software engineering.
Fourth, HMMs have certain inherent limitations, stemming from the central Markov assumption that the current state is ignorant of all states except for the previous one.
As reviewed in \cite{Ralph2016-kr}, this limits the scope of events that can be modeled using \partis, excluding correlation between different segments of the BCR~\cite{Volpe2008-xe,Elhanati2015-ld,Kidd2015-vt}, palindromic N-additions~\cite{Saada2007-kf}, complex strand interaction events~\cite{Kepler1996-kd,Jackson2007-ue}, or the appearance of tandem D segments~\cite{Larimore2012-lo}.
Some of these limitations could be avoided by using Conditional Random Fields (reviewed in \cite{MAL-013}), and although linear-chain conditional random fields enjoy many of the attractive computational properties of HMMs, this flexibility will come with a computational cost.
Fifth, \partis\ does not attempt to infer germline genotype, as do \cite{Gadala-Maria2015-uq}, and so treats genes and alleles on an equal footing.
We will treat this as a model-based inference problem in future development.
Sixth, we will continue to refine heuristics to provide the accuracy of the full likelihood-ratio calculation with minimal compute time.
We note, for instance, that a small decrease in the lower naive Hamming fraction threshold substantially improves performance for the seed \partis\ simulation compared to that shown here (in Fig~\ref{FIGseedCCFS}).

In additional future work, we will explore opportunities to combine clonal family inference and phylogenetics to obtain inference of complete B cell lineages.
This could potentially take the form of a phylo-HMM \cite{Siepel2005-mw}, although a more straightforward approach would be to take the product of a phylogenetic likelihood and a rearrangement likelihood \cite{Kepler2013-sy}.
For example, one might use HMM-based clustering as is described here with a high likelihood ratio cutoff to obtain a conservative collection of clusters, and then a phylogenetic criterion to direct further clustering.

In addition to these methodological improvements, we will also apply \partis\ to a variety of data sets for validation and to learn about the structure of natural repertoire.
For validation, there are some data sets, e.g. \cite{Wilson2000-bg}, which due to experimental setup have sequences known to make a clonal lineage.
Also, new microfluidics technology applied to BCR sequencing also gives heavy and light chain data \cite{DeKosky2013-iz,McDaniel2016-db}; although a single heavy chain clonal lineage can have light chains from independent rearrangement events, this type of data does provide further evidence of clonality for validation of clonal family inference procedures.
In addition to this sort of validation, there are now an abundance of data sets that can be used to characterize the size distribution of the clonal families in various immune states, such as health, immunization, and disease.

As a final note, \partis\ works to solve a challenging likelihood-based inference problem.
We recognize that in contrast to existing heuristic approaches based on sequence identity, our software is quite computationally demanding.
In this first paper we have developed the framework and overall approach, as well as many computational optimizations.
This optimization work is ongoing, and there remain many avenues for improvement.
As a comparison, likelihood-based phylogenetic inference has taken two decades of optimization to scale to tens of thousands of sequences at a time with approximate algorithms \cite{Price2010-fi}.
We are continually making improvements to the algorithm to make it scale to larger data sets and are committed to building algorithms that scale to the size of contemporary data sets.
Although such algorithms may end up being rather different than this version of \partis, we believe that likelihood-based algorithms will provide a solid foundation for large-scale molecular evolution studies of B cell maturation.

\section*{Methods}

\subsection*{Likelihood framework}
To introduce the way in which we use HMMs for BCR clustering, consider the canonical ``dishonest casino'' HMM \cite{Durbin1998-uq}.
In this introductory example, one imagines that a casino offers a game in which the casino alternates between a fair die and a die that is biased towards a given number, say 6.
Assume the dice are switched with probability $p$ each roll, corresponding to the HMM on two states, with a transition probability of $p$ between the states.
One favorite game of bioinformaticians is to infer the maximum likelihood identity of the die for each roll given a sampled sequence of roll outcomes, which is solved by the Viterbi algorithm.
The so-called forward algorithm, on the other hand, infers the marginal probability of a sequence of outcomes.

\begin{figure}[!ht]
\forarxiv{\includegraphics[width=3in]{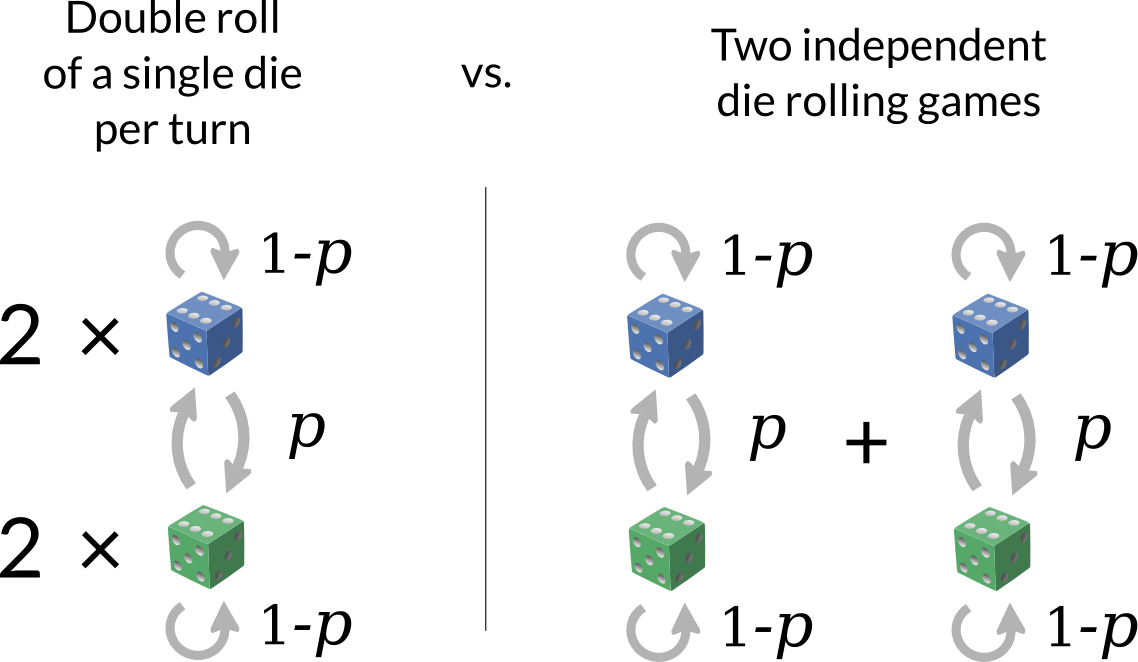}}
\caption{\
  {\bf Single and pair HMM likelihoods by analogy with a game of dice.}
}
\label{FIGdice}
\end{figure}

The likelihood ratio used in this paper fits into the metaphor with a slight variant of the game.
In this variant, a pair of outcomes (a.k.a. \emph{emissions}, in this case integers in the range 1 to 6) are sampled at each step.
The player knows that either the emissions came from rolling the same die twice and then switching out the die with probability $p$ after each step, or they came from rolling two dice which are independently switched out with probability $p$.
The new game, corresponding to the methods in this paper, is to figure out which of these scenarios is correct, and with what support.

The marginal probability of a sequence of emissions under the ``double roll'' scenario is that of a pair-HMM with transition probability $p$ with identical emission probabilities, while the latter ``two dice'' scenario is that of two independent HMMs.
The ratio of these two marginal probabilities is a likelihood ratio quantifying the strength of evidence for the ``double roll'' scenario.

Now, stepping back into the world of VDJ recombination, we will apply this logic to the HMM structure introduced in \cite{Ralph2016-kr}.
This HMM, building on prior work \cite{Volpe2005-uk,Gaeta2007-mz,Munshaw2010-mj}, has one state for each position in every V, D, and J gene, and a state for each of the joining N-regions for heavy chain sequences.
Light chain sequences are simpler, in that they have only V, J, and one N-region, and so for the rest of this methods section we will only describe the heavy chain procedure.

Continuing with the metaphor, the identity of the die (of which there are now many) for each roll corresponds either to an annotation of that nucleotide as being from a given non-templated insertion base, or as being from a specific nucleotide in a specific V, D, or J gene.
That is, a path through the HMM corresponds to an annotated ancestor sequence.
Our previous paper \cite{Ralph2016-kr} was focused on inferring these annotated ancestor sequences using the Viterbi algorithm.
Here we focus on the question of whether a group of sequences came from the same rearrangement event rather than on the annotated ancestor sequences themselves.
However, this distribution of annotated ancestor sequences is highly informative about the clonality of a group of sequences.
We would like use these annotated ancestor sequence inferences but avoid putting too much trust in one specific and necessarily uncertain inference, and instead account for the diversity for possible annotations.
We do so as follows.

Using $\sigma$ to designate paths and $x$ for a sequence, the marginal probability $\PP(x)$ of generating $x$ via any path is
\begin{equation*}
\PP(x) = \sum_{\sigma} \PP(x;\sigma),
\label{eq:margOne}
\end{equation*}
where $\PP(x;\sigma)$ designates the probability of generating $x$ with the path $\sigma$ through the HMM.
Now for a pair of sequences $x$ and $y$,
\begin{equation*}
\PP(x,y) = \sum_{\sigma} \PP(x,y;\sigma),
\label{eq:margTwo}
\end{equation*}
is the probability of generating both $x$ and $y$ using emissions from a single pass through the HMM.
Thus $\PP(x,y) / \left(\PP(x) \PP(y)\right)$ is a likelihood ratio such that values above 1 support the hypothesis that $x$ and $y$ come from the same rearrangement event and values less than 1 support the hypothesis that they do not.
Recall that all of these probabilities can be calculated efficiently via the forward algorithm.

More generally, if we would like to evaluate whether sequence sets $A$ and $B$ (each of which are assumed to descend from single rearrangement events) actually all came from a single rearrangement event.
For that we can calculate
\begin{equation}
\frac{\PP(A \cup B)}{\PP(A) \PP(B)}
\label{eq:ratio}
\end{equation}
where $\PP(X)$ can be calculated by a (simple) HMM if $X$ has one element, a pair-HMM if $X$ has two elements, etc., so in general a multi-HMM.
Note that this not a phylogenetic likelihood, but a rather strictly HMM-based likelihood, and so does not attempt to incorporate any tree structure into the computations.

\begin{figure}[!ht]
\forarxiv{\includegraphics[width=3.5in]{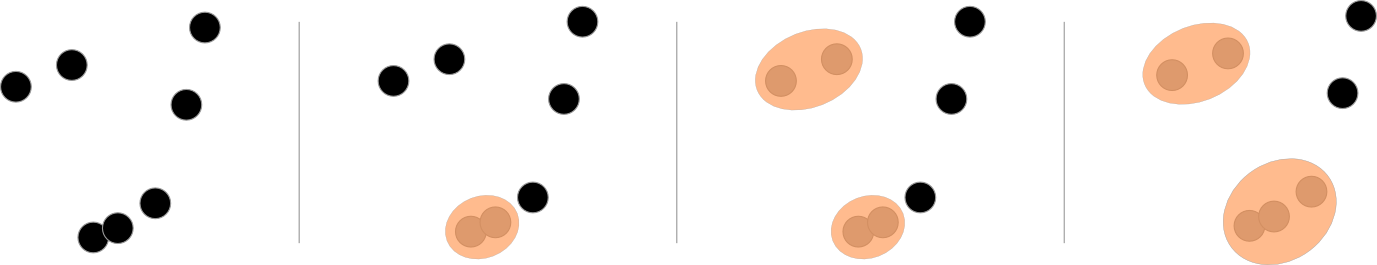}}
\caption{\
  {\bf Cartoon of agglomerative clustering.}
  Starting with singleton clusters, at each step we find the pair of clusters that maximizes \eqref{eq:ratio} and merge them.
}
\label{FIGagglomerative}
\end{figure}

We use this likelihood ratio for agglomerative clustering.
Specifically, at each step we pick the pair $A$ and $B$ that have the largest likelihood ratio \eqref{eq:ratio} and merge them by replacing $A$ and $B$ from the list of clusters and adding $A \cup B$ (Fig~\ref{FIGagglomerative}).
We stop agglomerating according to a likelihood ratio threshold, as described in the section after next.

\subsection*{Review of the HMM machinery of \cite{Ralph2016-kr}}
The HMM architecture we use is the same as that of \cite{Ralph2016-kr}, which for the most part follows previous work \cite{Volpe2005-uk,Gaeta2007-mz,Munshaw2010-mj} by representing each germline base in each V, D, and J allele as an HMM state.
All of these states can be combined to create a single HMM for the entire \vdj\ rearrangement process.
In order to allow likelihood contributions from the N-region, we replace the single insert state found in previous work with four states, corresponding to naive-sequence N-addition of A, C, G, and T.
The emissions of these four states are then treated as for actual germline states: the A state, for example, has a large probability of emitting an A, and a complementary probability (equal to the observed mutation probability) of emitting one of the other three bases.

Our application of HMMs also differs from previous work using HMMs for B cell receptor sequence analysis in that we do inference under a model which simultaneously emits an arbitrary number of symbols $k$.
When $k=2$ this is typically called a pair HMM \cite{Durbin1998-uq}, and we call the generalized form a multi-HMM ($k \geq 2$).
One can also think of this as doing inference while constraining all of the sequences to come from the same path through the hidden states of the HMM.
In our setting, the $k$ sequences resulting from such a multi-HMM model are the various sequences deriving from a single rearrangement event (which differ only according to point substitution from somatic hypermutation).
HMM inference is performed by an efficient new HMM compiler, called \ham, which we wrote to inference on an arbitrary (multi-)HMM specified via a simple text file (\url{https://github.com/psathyrella/ham/}).

\subsection*{Naive Hamming fraction thresholds}
A straightforward application of hierarchical clustering in this setting, in which the likelihood ratio is computed for every cluster at every stage of the algorithm, would not scale to more than a few hundred sequences.
Thus as described above, we also use Hamming fraction (Hamming distance divided by sequence length) between inferred naive sequences to avoid expensive likelihood ratio computation.
In order to compare unequal-length sequences, we first align the conserved cysteine in every sequence, and then pad all sequences on both ends with ambiguous nucleotides until they are all the same length.
In addition to point \partis\, described as an approximate method above, we also use naive Hamming fraction in the full \partis\ method in order to identify sequences that are either very likely or very unlikely to be clones.
We assume that clusters which differ by more than 0.08 in naive Hamming fraction are not clonal, and therefore avoid calculating the full likelihood for these cases.
This threshold is for repertoires with typical mutation levels (around 5\%); we find that increasing the threshold as mutation increases (to 0.15 at 20\% mutation) provides optimal performance.
We interpolate and extrapolate linearly for other mutation levels.
In addition, we assume that clusters that are closer than 0.015 (regardless of mutation levels) in naive Hamming fraction are clonal, and merge these without calculating the full likelihood.
While the naive Hamming fraction only takes into account the Viterbi path (i.e.\ it does not sum over all potential annotated ancestor sequences), and it has no probabilistic interpretation, it has the not insignificant virtue of being much faster to calculate.

\subsection*{Likelihood ratio thresholds}
According to standard statistical theory, we should merge an \emph{a priori} specified pair of clusters $A$ and $B$ when the likelihood ratio \eqref{eq:ratio} is greater than one.
However, in the midst of a series of agglomerations, we are not in the setting of a single decision for clusters that have been presented to us.
Instead, at every stage we are comparing a quadratic number of potential merges and asking if the pair of clusters with the \emph{largest} likelihood ratio deserve to be merged.
This effectively presents substantial multiple testing issues: even when no more clusters should be merged, the nonzero-width of the empirical likelihood ratio distribution will typically have points above one.
Furthermore, the marginal probability $\PP(A)$ of, say, the $k$th largest cluster after some number of merges is going to be biased by the fact that the sequences in that cluster were selected to merge.
Such issues are not new in computational biology \cite{Goldman2000-qw}.
We also note that we are only calculating this likelihood ratio when pairs of sequences are similar enough in their inferred naive sequences to merit such a likelihood ratio calculation, further taking us from the statistically ideal setting.

\begin{table}[!ht]
{\small
  \centering
\begin{tabular}{cc}
  \textbf{\pbox{20cm}{candidate \\ cluster size}\vspace{3pt}} & \pbox{20cm}{\textbf{log ratio}\\ \textbf{threshold}} \\
\hline
2 & 18 \\
3 & 16 \\
4 & 15 \\
5 & 14 \\
$\ge 6$ &  13 \\
\hline
\end{tabular}
\caption{\
  {\bf Log likelihood ratio merging threshold as a function of \ccs.}
}\label{TABLElratioThreshold}}
\end{table}

We have found it useful to use a likelihood ratio threshold greater than 1, and use a threshold that decreases as the \emph{candidate cluster size}, i.e. the size of a proposed cluster, increases (Table~\ref{TABLElratioThreshold}).
These values were selected as a trade-off between accurate reconstruction of large clonal families on the one hand, and accuracy at the low end of the cluster size distribution on the other.
Thus if we want to minimize the chance of missing highly-mutated members of a large clonal family we should choose lower thresholds, but if we instead want to avoid mistakenly merging unrelated singletons we should choose higher ones.
In light of this, the thresholds can be set on the command line.

\subsection*{Insertion and deletion mutations}
While it would be straightforward in principle to account for insertions and deletions (\emph{indels}) during somatic hypermutation within the HMM by adding extra transitions for deletions and extra states for insertion, this approach would entail a very substantial computational cost.
When restricting to substitution mutations, each germline state can either transition to the next germline state, or it can leave the region.
If we allowed indels within the V, D, and J segments, however, each state would also need to investigate the probability to transition to a special insertion state as well as to any subsequent germline state.
This would introduce a quadratic dependence on the number of states and the resulting algorithm would not be able to analyze realistically-sized data sets.

We thus instead adopt an approach to indel mutations based on the annotation from our preliminary Smith-Waterman step (implemented with \ighutil~\cite{McCoy2015-qi}).
In cases where \ighutil\ detects an insertion with respect to a germline segment, we ``reverse'' the insertion by removing it from the query sequence.
Similarly, candidate deletions are reversed by inserting the corresponding germline bases from the best germline match when the putative deletion happens in a germline segment.
In both cases the original sequences are maintained, but the \partis\ processing of the sequences is done on the modified sequences.

As with any Smith-Waterman implementation, this approach depends on several arbitrary parameters: the match and mismatch scores and the gap-opening penalty.
In particular, a larger gap-opening penalty relative to the match/mismatch scores decreases sensitivity to indels.
On all samples which we have encountered, a good initial set of match:mismatch scores is 5:1.
Sequences with lower mutation rates, for which 5:1 is less optimal, are returned with no D segment match, and then re-run with match:mismatch scores of 5:2.
Sequences which still have no D matches are then rerun with scores of 5:3.
This procedure gives good results in all parameter regimes which we have encountered in the data.
Similarly, we find that a gap-opening penalty of 30 provides good sensitivity to indel mutations in simulation.
Each of these parameters may also be set with a command line flag. 


In order to test the effectiveness of this method, we made simulated samples in which each sequence has a 50\% chance of having an indel mutation after being generated on a tree.
Each indel has equal probability of being an insertion or a deletion, and the indel's position is chosen from the uniform distribution either on the bulk of the V segment (between position 10 and the conserved cysteine), or on the CDR3.
The length of each indel is drawn from a geometric distribution with mean 5.
These samples are not intended to mimic any particular data set, but are instead designed to provide an extremely stringent test of performance in the presence of indel mutations (Fig~\ref{FIGmetricsIndels}).

\subsection*{Approximate Methods}
The accuracy of the full likelihood framework which we have described above does not come without some computational cost.
As such we have also implemented two other algorithms which make some reasonable trade-offs in accuracy in order to gain some speed.

\subsubsection*{Point \partis}
One of the biggest contributors to both annotation and partitioning accuracy comes from our multi-HMM framework's ability to run simultaneously on an arbitrary number of sequences.
Since this ability is entirely separate to the summation over all possible rearrangements, it makes sense to decouple the two in order to optimize for speed.
We can, in other words, cluster using the single best (Viterbi) annotated ancestor sequence for all sequences in a cluster (inferred simultaneously on the whole cluster with the multi-HMM), without summing over all germline genes and all rearrangement boundaries.
We call this \emph{point \partis}, to emphasize that it uses the best point (i.e.\ single) annotation inference to do clustering.
In order to cluster on these inferred naive sequences, we use the hierarchical agglomeration described above, but with Hamming fraction as the metric (instead of log likelihood ratio).
As in the case of the likelihood ratio merging thresholds described above, we perform a simple optimization procedure on a wide variety of simulation samples which span the range of possible lineage structures and mutation levels that we observe in real data.
For typical (low) mutation levels near 5\%, we use a threshold of 0.035; the threshold then increases to 0.06 as the mutation frequency reaches 20\%.
Simple linear interpolation (extrapolation) is used inside (outside) of this range.
Note that these thresholds are much tighter than those mentioned above for full \partis\ optimization: while above we were trying to exclude cases where there was any doubt as to their clonality, here we are attempting to accurately divide clonal from non-clonal clusters in the naive Hamming distribution.
Comparing to Fig 7 in~\cite{Ralph2016-kr}, we note that this threshold is equivalent to the expected fractional error in the inferred naive sequence.

\subsubsection*{\vsearch\ \partis}
The \emph{point} method, however, still performs full all-vs-all comparisons on the entire data set, and recalculates the full Viterbi naive sequence on each cluster each time more sequences are added.
While this is a good way to ensure the best accuracy, there exist clustering algorithms with many optimizations which trade some of this accuracy for improved speed.
\vsearch~\cite{vsearch} is one such tool, and we have included a version of \partis\ which infers the Viterbi naive sequence for each single query, and then passes these sequences to \vsearch.
This sacrifices some accuracy, particularly on larger clonal families, but is extremely fast.
We use \vsearch\ version 1.1.3 in \texttt{cluster fast} mode with the maximum accept and reject thresholds set to zero, and the \texttt{id} threshold set (again, based on coarse heuristic optimization) to one-half the threshold described above for point \partis.

\subsection*{Reconstruction of selected lineages}
We have added an option to reconstruct the lineage of a user-specified sequence using full \partis, for situations in which one is only interested in one specific clonal family.
We call such a user-specified sequence a \emph{seed} sequence.
This is shown as ``full \partis\ (seed)'' (Figs~\ref{FIGsimilarityMatricesSimu50LeavesMut1} and~\ref{FIGsimilarityMatricesSimu50LeavesMut4}).
Here we chose a seed sequence at random from a randomly-selected ``large'' cluster, where ``large'' means with size greater than or equal to the mean $N$ leaves for the sample.
It can be seen that this method accurately reconstructs the single lineage of interest while running much more quickly than the other methods (Fig~\ref{FIGtimeRequired}).

\subsection*{Simulation and validation}
\begin{figure}[!ht]
  \forarxiv{\includegraphics[width=5.5in]{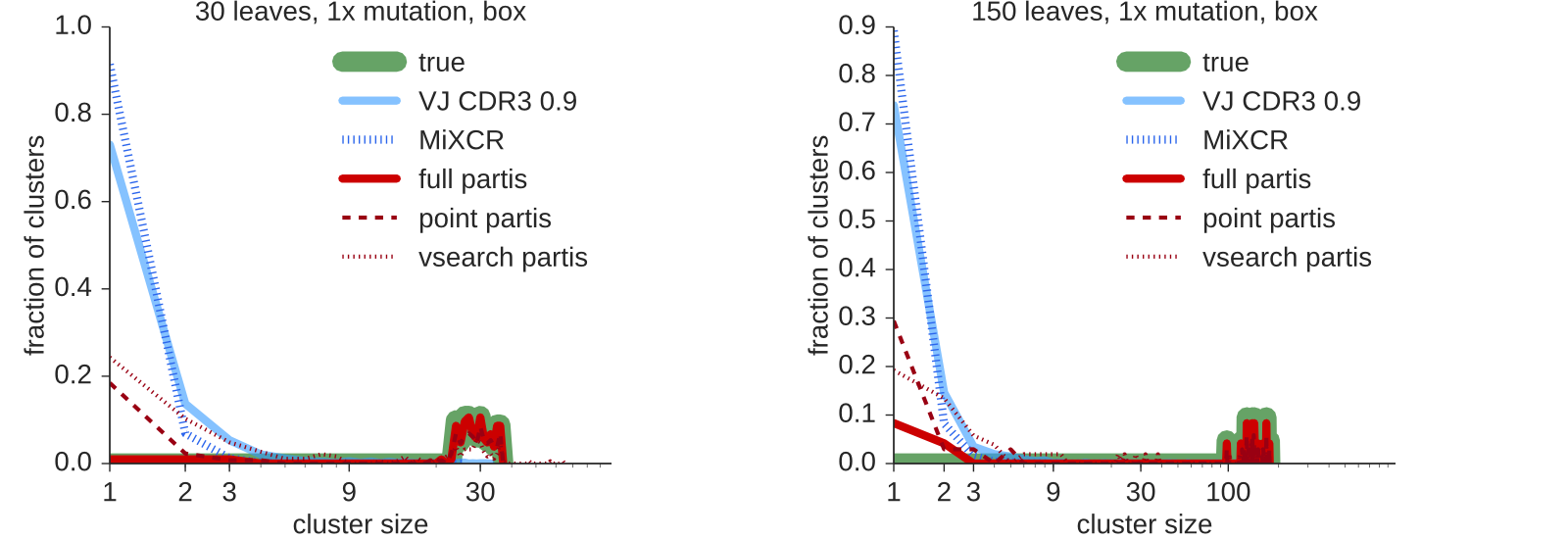}}
\caption{\
  {\bf True and inferred cluster size distributions for simulation with non-geometric cluster size distributions.}
  Results are shown for box distributions with mean 30 (left) and 150 (right).
}
\label{FIGclusterSizeBox}
\end{figure}

\begin{figure}[!ht]
  \forarxiv{\includegraphics[width=5.5in]{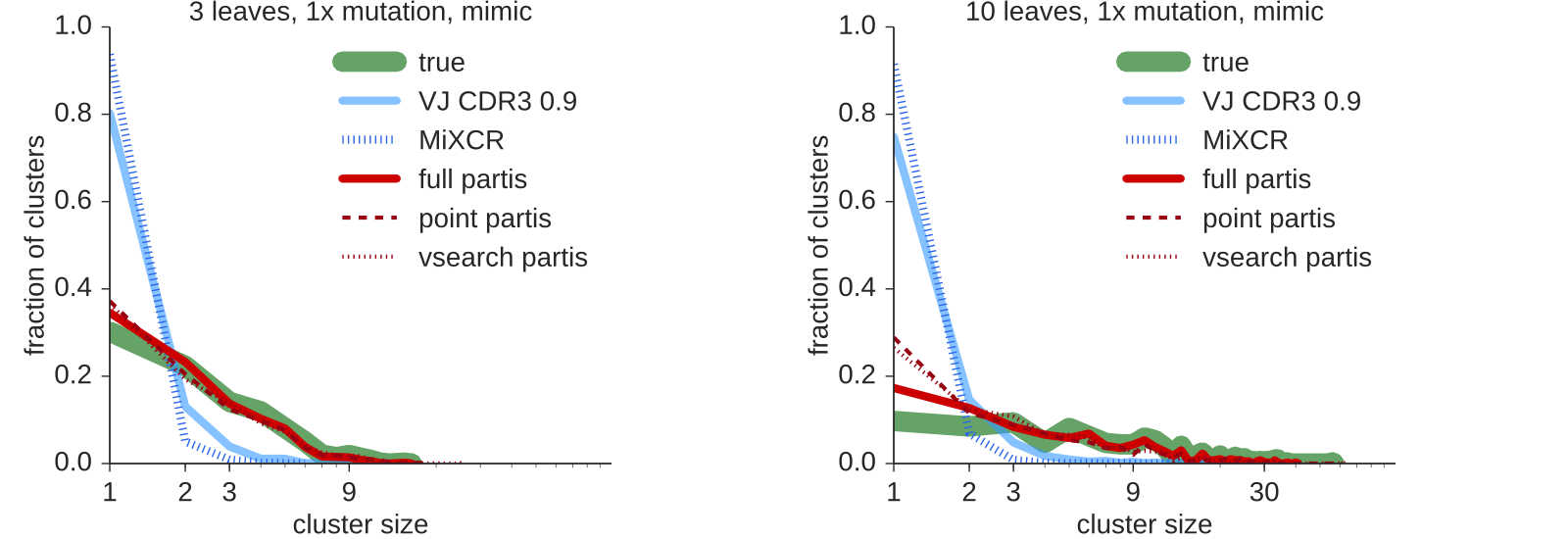}}
\caption{\
  {\bf True and inferred cluster size distributions for simulation with shorter read lengths.}
  Results are shown for samples with reads of length 130 bases centered on the CDR3 (which mimics the Adaptive data set), for geometric cluster size distributions with mean 3 (left) and 10 (right).
}
\label{FIGclusterSizeMimic}
\end{figure}

To benchmark results, we simulate sequences using the procedure described in~\cite{Ralph2016-kr}.
This provides a bountiful supply of sequences for which the correct lineage structures are known, and with any desired combination of tree topologies and mutation parameters, but with all other properties mimicking empirical values.
Briefly, the simulation proceeds by sampling a set of parameters defining a single rearrangement (e.g. V exonuclease deletion length, V allele, etc.) from their empirical joint distribution observed in a data set.
Then TreeSim \cite{Stadler2011-sz} is used to simulate a tree and Bio++ \cite{bppseqgen} is used to simulate sequences.
We emphasize that these sequences are not generated at any stage using \partis' HMM, and no information concerning the simulation is fed to the clustering code other than the simulated sequences.
The number of leaves (BCR sequences per clonal family) is distributed geometrically with the indicated mean value in all figures except Figs~\ref{FIGseedCCFS} and~\ref{FIGclusterSizeBox}.
In Fig~\ref{FIGseedCCFS} we have used a Zipf (power law) distribution.
In Fig~\ref{FIGclusterSizeBox}, on the other hand, we have used a box-shaped distribution to check that our methods do not depend on a monotonically decreasing distribution.
In order to simulate a given number of sequences, we simply divide the desired number of sequences by the expected number of sequences per clone and simulate the resulting number of clones.
For indel simulations, half of the simulated sequences have a single indel, whose length is drawn from a geometric distribution with mean 5.
In order to emphasize the importance of the indel's location, we show samples where they are distributed evenly either within the CDR3, or within the bulk of the V segment (specifically between position 10 and the conserved cysteine).

\begin{figure}[!ht]
  \forarxiv{\includegraphics[width=5.5in]{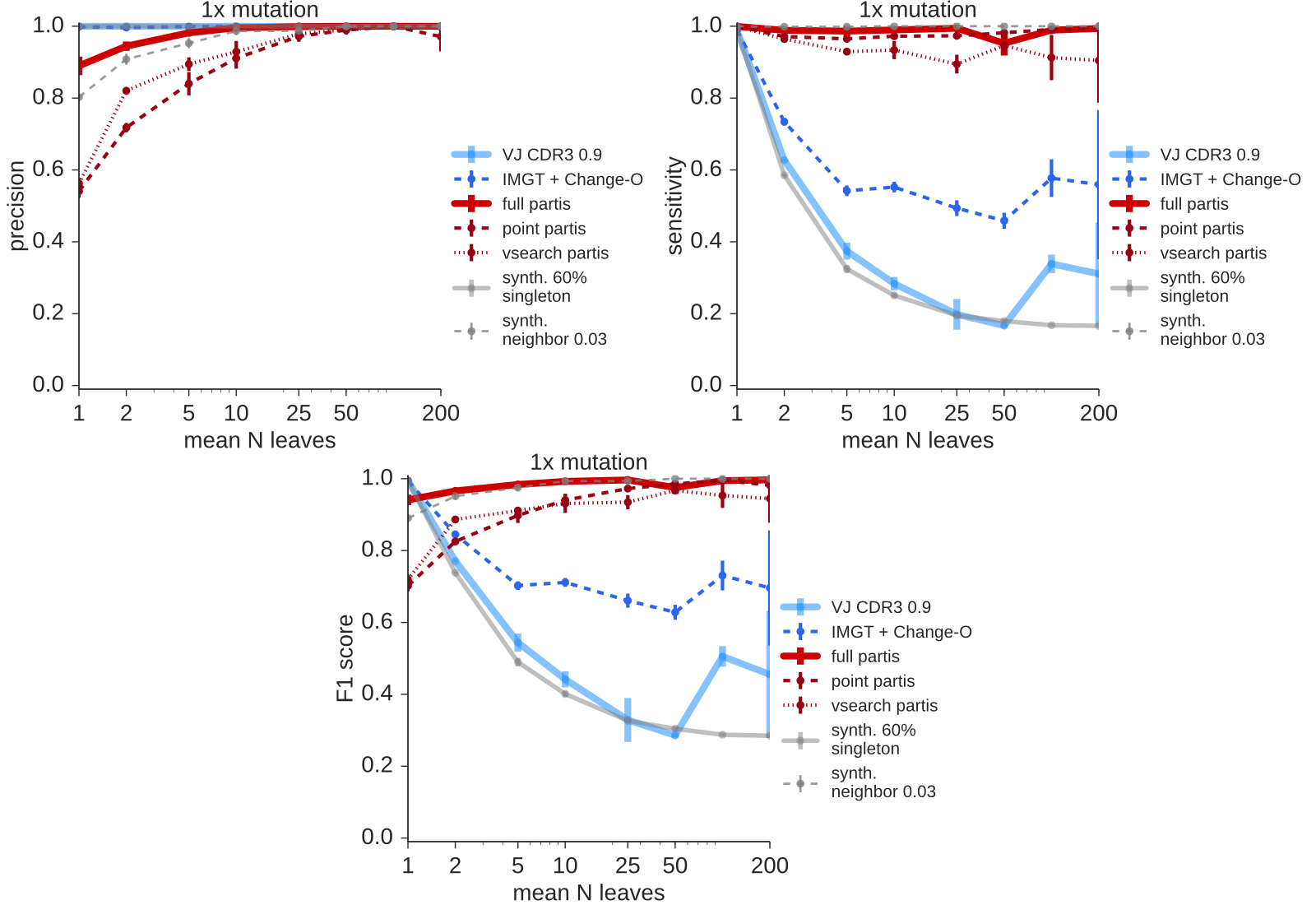}}
\caption{\
  {\bf Similarity between inferred and true partitions for the various clustering methods at typical ($1 \times$) mutation levels}
  \ccfFigExplain
  These plots also include \emph{synthetic} partitions, which for purposes of comparison generate incorrect partitions starting from the true partition (``synth.'', see text for details).
}\label{FIGccfMut1}
\end{figure}

\begin{figure}[!ht]
  \forarxiv{\includegraphics[width=5.5in]{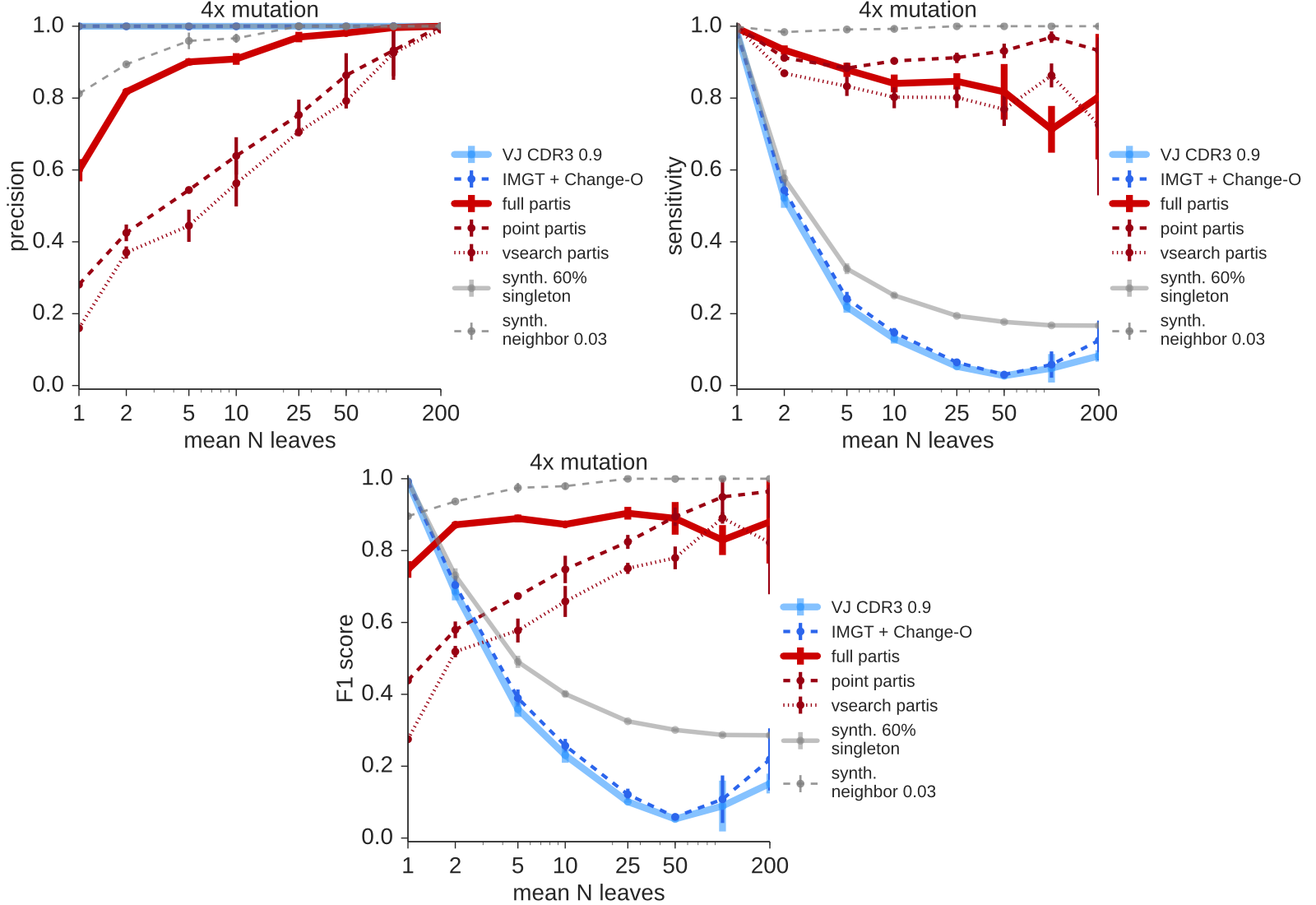}}
\caption{\
  {\bf Similarity between inferred and true partitions for the various clustering methods at high ($4 \times$) mutation levels} with the same labels as in Figure~\ref{FIGccfMut1}.
}\label{FIGccfMut4}
\end{figure}

We use per-sequence averages of sensitivity and precision to quantify clustering accuracy.
In this context, a true positive (TP) statement about a sequence $x$ is the correct identification of another sequence in $x$'s clonal family, i.e. correctly clustering a sequence with $x$.
A false postive (FP) statement is incorrectly clustering a sequence with $x$, while a false negative (FN) statement is not clustering a sequence with $x$ that should be clustered.
\[
\sensitivity_x = \frac{|\TP_x|}{|\TP_x + \FN_x|} \qquad
\precision_x = \frac{|\TP_x|}{|\TP_x + \FP_x|}
\]
Thus, as described above, the precision for a given read is the fraction of sequences in its inferred cluster to which it is truly clonally related.
The sensitivity for a given read is the fraction of sequences in its true cluster that appear in its inferred cluster.
We average these two quantities over all sequences (Figs~\ref{FIGccfMut1NoSynthetic},~\ref{FIGccfMut1}, and~\ref{FIGccfMut4}).
These figures also show the average harmonic mean of this sensitivity and precision (a.k.a. F1 score), as an aggregate measure of the quality of the clustering.

We also show \emph{intersection matrices}: the matrix of intersection sizes between pairs of large clusters in two partitions (examples in Figs~\ref{FIGsimilarityMatricesSimu50LeavesMut1}, and~\ref{FIGsimilarityMatricesSimu50LeavesMut4},~\ref{FIGadaptive}, and~\ref{FIGstanford}; the full set of plots is available at \supplementUrl.
To make these plots, we first take the \nbiggestclusters\ largest clusters from each of the two partitions.
Each non-white square indicates that there was a non-empty intersection between the two clusters; the square is shaded by the size of the clusters' intersection divided by their mean size.
The position of the square shows the relative sizes of the two clusters.
Thus a value of 1.0 implies identity, so very similar partitions will show many dark squares near the diagonal, and will also have similar cluster sizes marked on the $x$ and $y$ axes.

\subsubsection*{Performance versus sample size}
Given the large size of modern deep sequencing data sets, we have also investigated performance as a function of sample size.
This function depends on the clonal lineage structure.
At one extreme, a sample with only a few sequences stemming from a few clonal families is generally trivial to partition even just by visual inspection.
As the number of clonal families increases, however, each family becomes closer and closer to other families, and it becomes more and more difficult to distinguish between them.
At the point where the naive sequences corresponding to each family are separated by only a few bases, accurate overall clustering becomes impossible even in principle, since a difference of only a few bases which stems from rearrangement cannot be distinguished from somatic hypermutation.

\begin{figure}[!ht]
  \forarxiv{\includegraphics[width=5.5in]{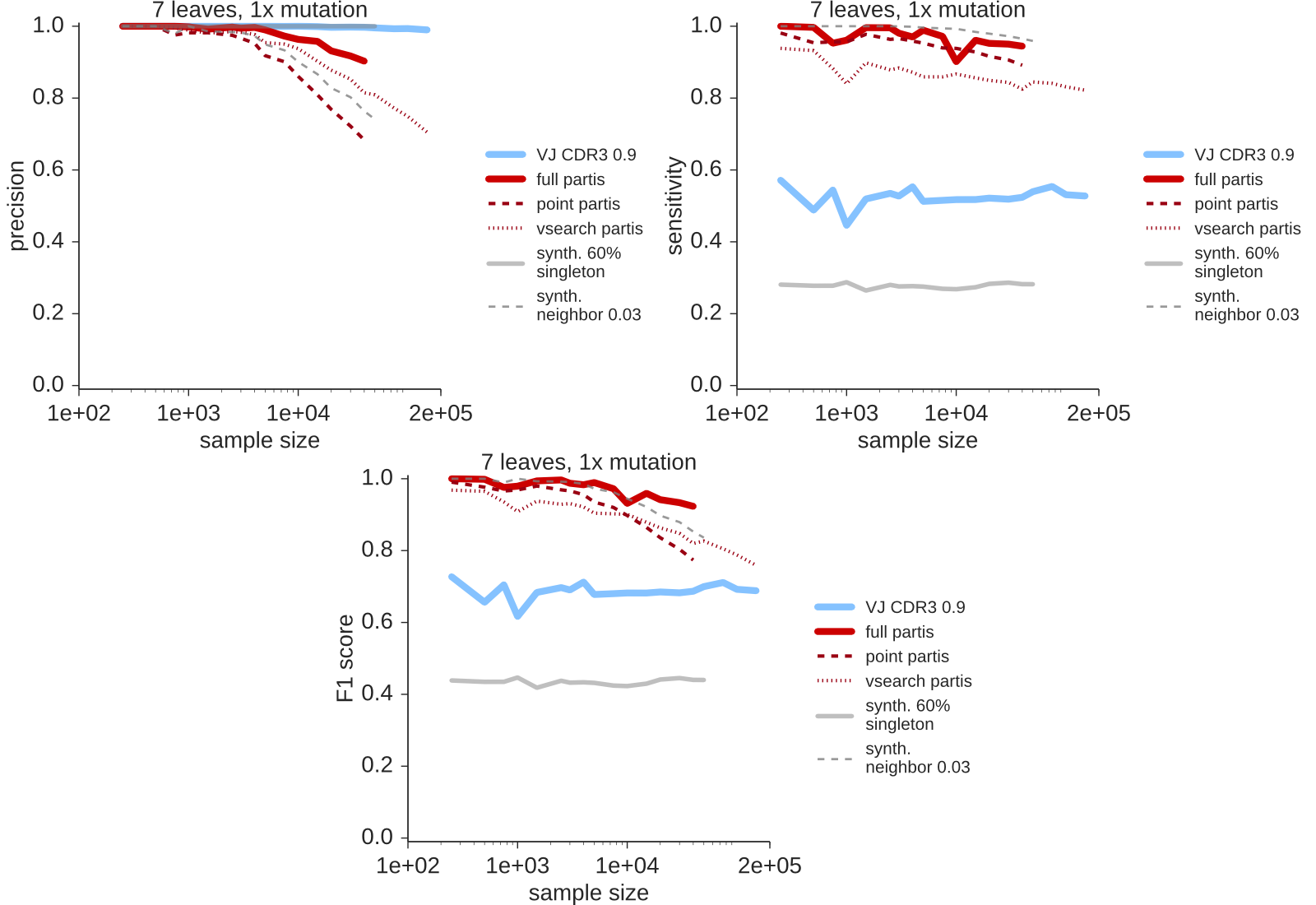}}
\caption{\
  {\bf Overall clustering quality metrics as a function of sample size, on simulation corresponding to patient 15-12 in the Vollmers data set.}
  See Figure~\ref{FIGccfMut1} for figure labels.
}\label{FIGmetricsVsSampleSize}
\end{figure}

\begin{figure}[!ht]
  \forarxiv{\includegraphics[width=5.5in]{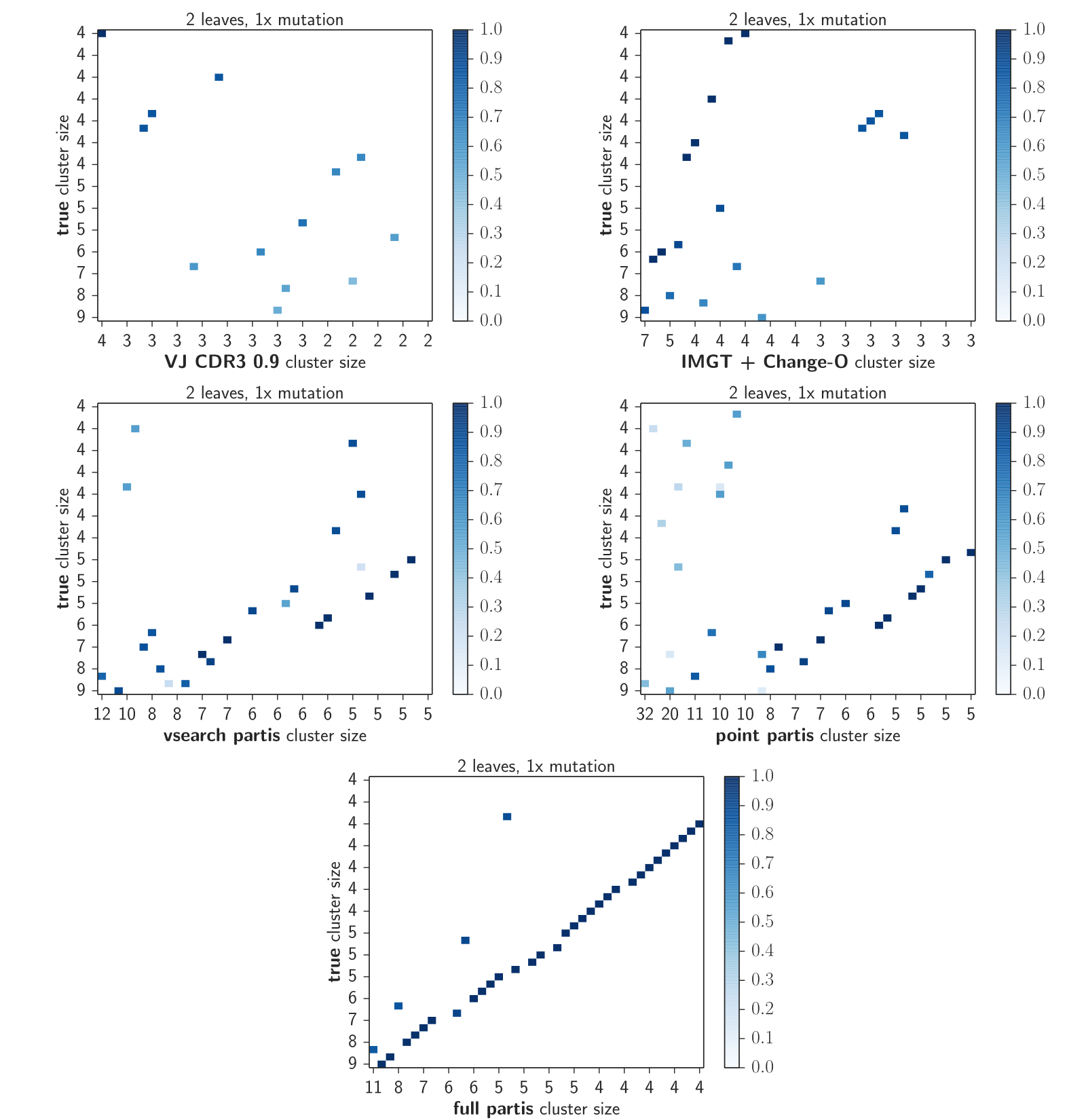}}
\caption{\
  {\bf Fraction of sequences per cluster in common with the true partition on simulation with a mean 2 geometric distribution for the number of leaves at typical ($1 \times$) mutation levels.}
\similarityMatrixExplain
}
\label{FIGsimilarityMatricesSimu2LeavesMut1}
\end{figure}

\begin{figure}[!ht]
  \forarxiv{\includegraphics[width=5.5in]{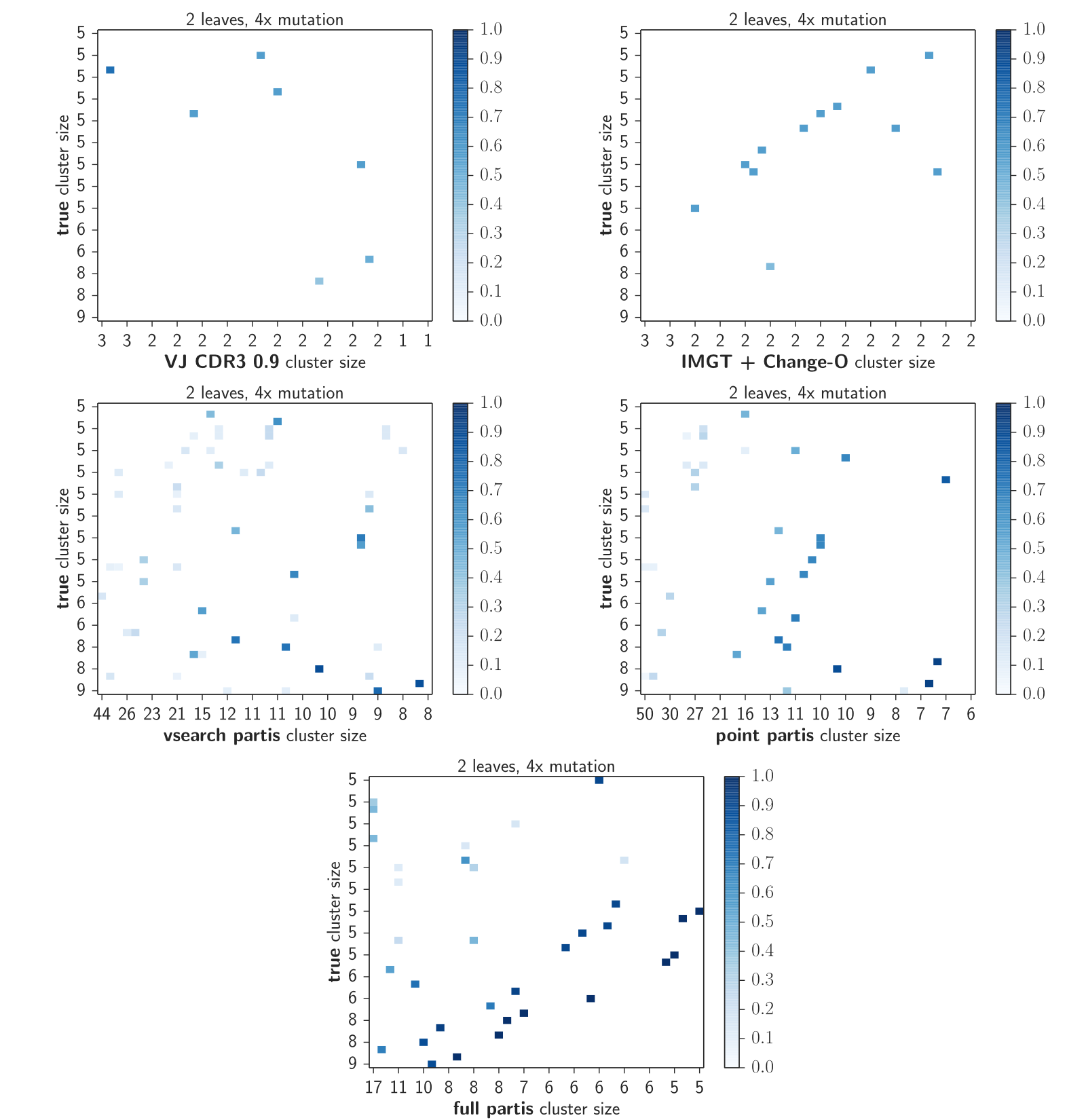}}
\caption{\
  {\bf Fraction of sequences per cluster in common with the true partition on simulation with a mean 2 geometric distribution for the number of leaves at high ($4 \times$) mutation levels.}
  \similarityMatrixExplain
}
\label{FIGsimilarityMatricesSimu2LeavesMut4}
\end{figure}

\begin{figure}[!ht]
  \forarxiv{\includegraphics[width=5.5in]{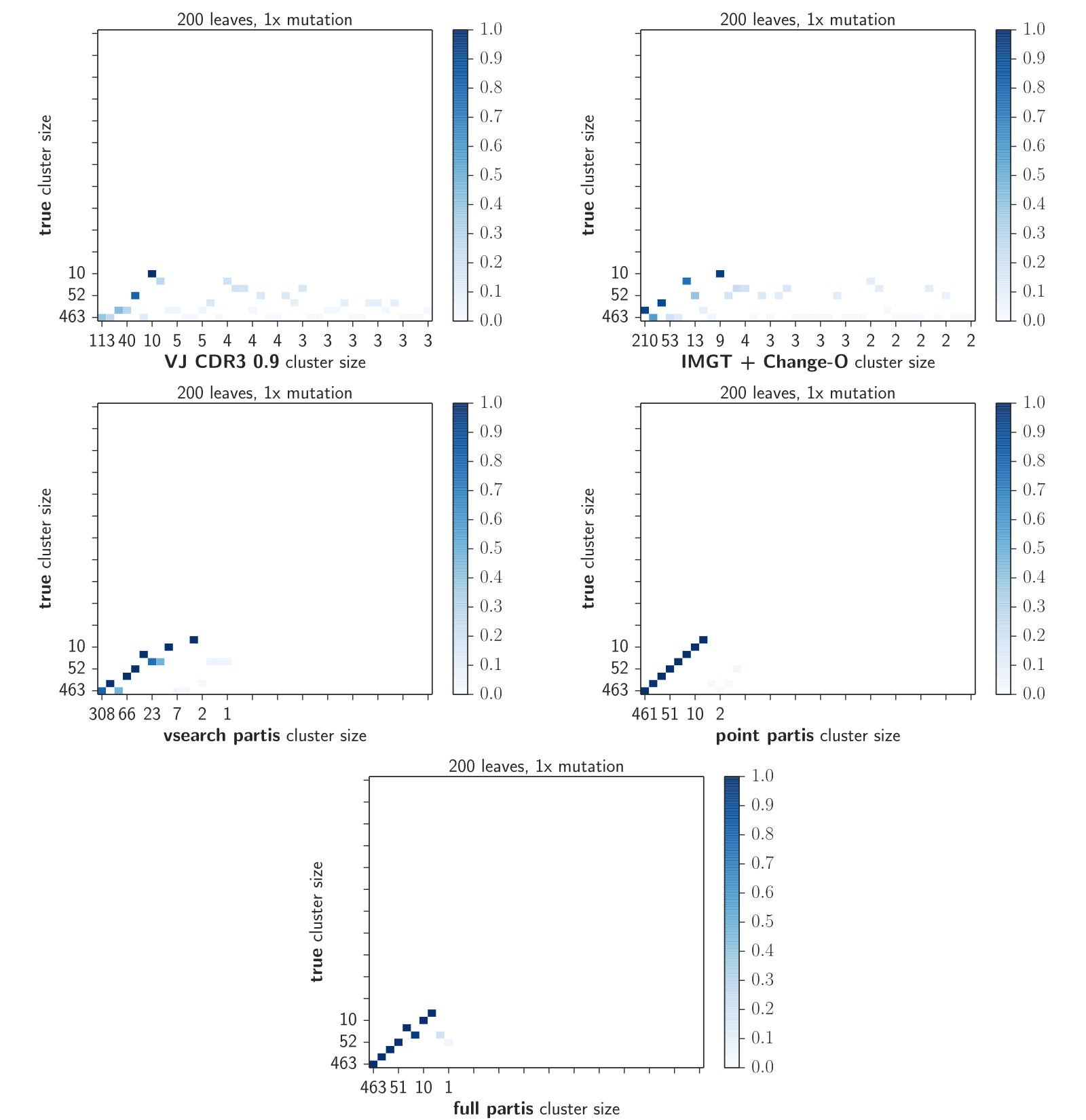}}
\caption{\
  {\bf Fraction of sequences per cluster in common with the true partition on simulation with a mean 200 geometric distribution for the number of leaves at typical ($1 \times$) mutation levels.}
  \similarityMatrixExplain
}
\label{FIGsimilarityMatricesSimu200LeavesMut1}
\end{figure}

\begin{figure}[!ht]
  \forarxiv{\includegraphics[width=5.5in]{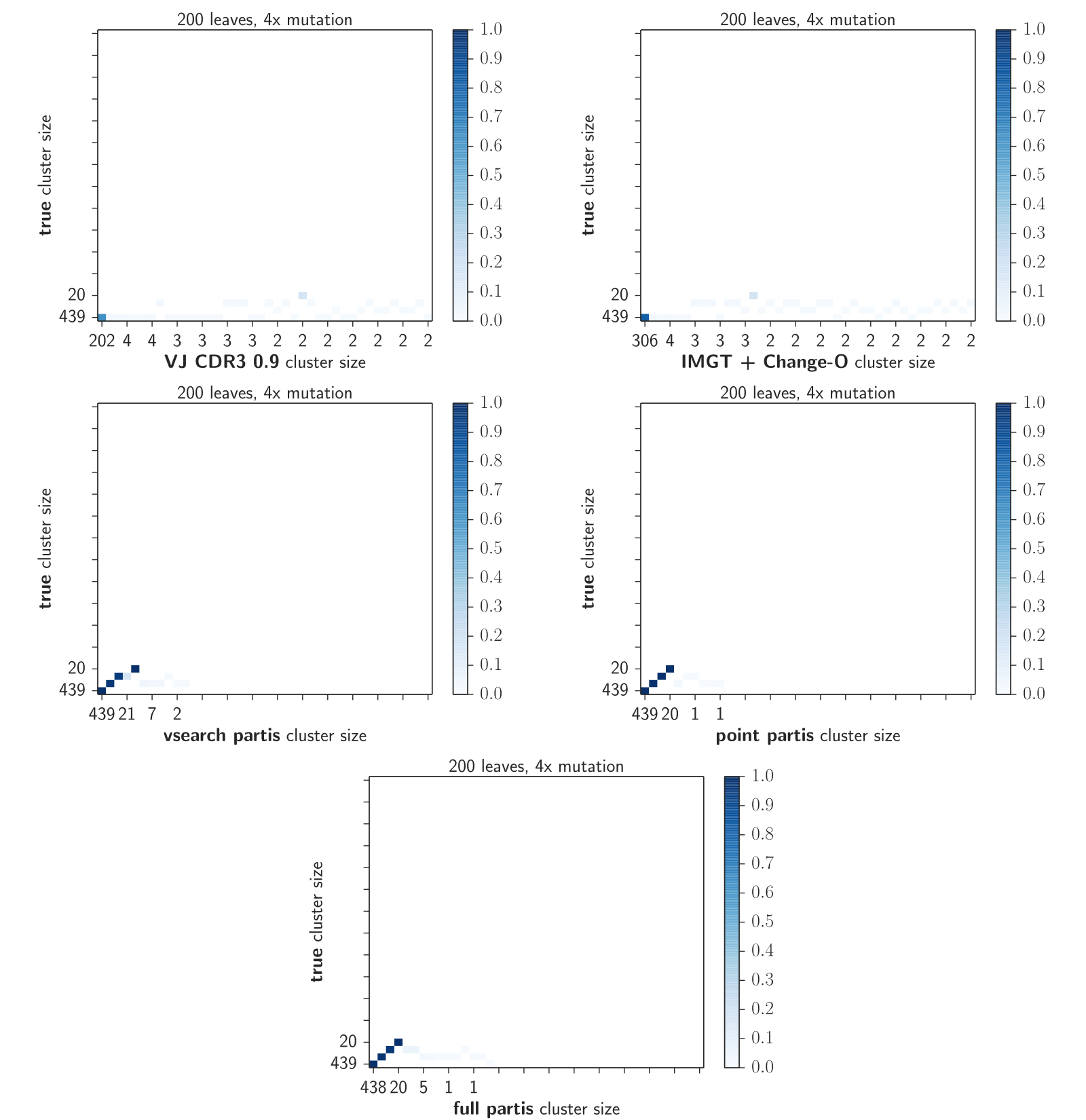}}
\caption{\
  {\bf Fraction of sequences per cluster in common with the true partition on simulation with a mean 200 geometric distribution for the number of leaves at high ($4 \times$) mutation levels.}
  Results are shown for the simulation sample with a mean 200 geometric distribution for the number of leaves.
  \similarityMatrixExplain
}
\label{FIGsimilarityMatricesSimu200LeavesMut4}
\end{figure}

In order to evaluate this performance we show several performance metrics as a function of sample size (Fig~\ref{FIGmetricsVsSampleSize}).
Here we show the two complementary precision and sensitivity metrics in the top row, and their harmonic mean (F1 score) in the bottom row.
It can be seen the behavior of the \partis\ with sample size is similar to that of the synthetic partition which joins neighboring true clusters which are closer than some threshold.
This is expected, and demonstrates that performance of the \partis\ method decreases as the number of true naive rearrangements in the sample increases, and thus the clonal family inference problem is becoming inherently more difficult.

\subsection*{Parallelization}

Non-independence of clustering steps poses a challenge for parallelization, and we approach this challenge with a combination of principled probability calculations and reasonable heuristics.
The basic strategy is to begin with a large number of processes, each running on a small subset of the data sample.
When each of these processes finishes clustering its allotted sequences, it reports back to the parent program, which collects the results from each subprocess and reapportions the resulting clusters among a new, smaller number of processes for the next step.
The process then repeats until we arrive at a single process which is comparing all clusters against all other clusters.
On the face of it, each step in this scheme would take much longer than the previous one since it is comparing more sequences.
However, because each process caches all the likelihoods it calculates, and because both factors in the denominator for each likelihood ratio~\eqref{eq:ratio} is guaranteed to have been calculated in a previous step, we can choose the process number reduction ratio such that each stage of paralellization takes roughly the same time.

An important part of this process is the allotment of sequences to processors.
At present we apportion them randomly in order to achieve a (very) roughly equal number of computations per process.
This is far from ideal, however, because we want to merge clonal sequences as soon as possible in order to avoid unnecessary comparisons to non-clonal sequences.
This must be balanced, however, by the need to evenly distribute the workload across all processes.
In the future we will study in more detail the optimal allotment scheme, and anticipate substantial speed increases.

\section*{Acknowledgements}
The authors would like to thank
Vu Dinh,
Julie Overbaugh,
Vladimir Minin,
and
Chaim Schramm
for support and helpful discussions,
and the three anonymous peer reviewers.
Jakub Otwinowski gave excellent feedback on software usability.
This research supported by National Institutes of Health R01 GM113246 (PI Matsen), R01 AI103981, and U19 AI117891.

\notforarxiv{
  \section*{Figure Legends}
}

\bibliographystyle{plos2015}
\bibliography{hmm-clustering-paper}

\begin{thebibliography}{10}

\bibitem{Melchers2015-co}
Melchers F.
\newblock Checkpoints that control {B} cell development.
\newblock J Clin Invest. 2015 Jun;125(6):2203--2210.
\newblock Available from: \url{http://dx.doi.org/10.1172/JCI78083}.

\bibitem{Victora2014-hm}
Victora GD, Mesin L.
\newblock Clonal and cellular dynamics in germinal centers.
\newblock Curr Opin Immunol. 2014 Jun;28:90--96.
\newblock Available from: \url{http://dx.doi.org/10.1016/j.coi.2014.02.010}.

\bibitem{Eisen1964-ls}
Eisen HN, Siskind GW.
\newblock Variations in affinities of antibodies during the immune response.
\newblock Biochemistry. 1964 Jul;3:996--1008.
\newblock Available from: \url{http://www.ncbi.nlm.nih.gov/pubmed/14214095}.

\bibitem{Cooper2015-pk}
Cooper MD.
\newblock The early history of {B} cells.
\newblock Nat Rev Immunol. 2015 Mar;15(3):191--197.
\newblock Available from: \url{http://dx.doi.org/10.1038/nri3801}.

\bibitem{Doria-Rose2014-vi}
Doria-Rose NA, Schramm CA, Gorman J, Moore PL, Bhiman JN, DeKosky BJ, et~al.
\newblock Developmental pathway for potent {V1V2-directed} {HIV-neutralizing}
  antibodies.
\newblock Nature. 2014 2~Mar;Available from:
  \url{http://dx.doi.org/10.1038/nature13036}.

\bibitem{Wu2015-uw}
Wu X, Zhang Z, Schramm CA, Joyce MG, Do~Kwon Y, Zhou T, et~al.
\newblock Maturation and Diversity of the {VRC01-Antibody} Lineage over 15
  Years of Chronic {HIV-1} Infection.
\newblock Cell. 2015 8~Apr;Available from:
  \url{http://dx.doi.org/10.1016/j.cell.2015.03.004}.

\bibitem{Mascola2013-mt}
Mascola JR, Haynes BF.
\newblock {HIV-1} neutralizing antibodies: understanding nature's pathways.
\newblock Immunol Rev. 2013 Jul;254(1):225--244.
\newblock Available from: \url{http://dx.doi.org/10.1111/imr.12075}.

\bibitem{Dosenovic2015-ek}
Dosenovic P, von Boehmer L, Escolano A, Jardine J, Freund NT, Gitlin AD, et~al.
\newblock Immunization for {HIV-1} Broadly Neutralizing Antibodies in Human Ig
  Knockin Mice.
\newblock Cell. 2015 18~Jun;161(7):1505--1515.
\newblock Available from: \url{http://dx.doi.org/10.1016/j.cell.2015.06.003}.

\bibitem{Bashford-Rogers2013-xv}
Bashford-Rogers RJM, Palser AL, Huntly BJ, Rance R, Vassiliou GS, Follows GA,
  et~al.
\newblock Network properties derived from deep sequencing of human {B}-cell
  receptor repertoires delineate {B}-cell populations.
\newblock Genome Res. 2013 Nov;23(11):1874--1884.

\bibitem{Volpe2005-uk}
Volpe JM, Cowell LG, Kepler TB.
\newblock {SoDA}: implementation of a {3D} alignment algorithm for inference of
  antigen receptor recombinations.
\newblock Bioinformatics. 2005 15~Dec;22(4):438--444.
\newblock Available from:
  \url{http://dx.doi.org/10.1093/bioinformatics/btk004}.

\bibitem{Gaeta2007-mz}
Ga{\"{e}}ta BA, Malming HR, Jackson KJL, Bain ME, Wilson P, Collins AM.
\newblock {iHMMune-align}: hidden {M}arkov model-based alignment and
  identification of germline genes in rearranged immunoglobulin gene sequences.
\newblock Bioinformatics. 2007 26~Apr;23(13):1580--1587.
\newblock Available from:
  \url{http://dx.doi.org/10.1093/bioinformatics/btm147}.

\bibitem{Munshaw2010-mj}
Munshaw S, Kepler TB.
\newblock {SoDA2}: a Hidden {M}arkov Model approach for identification of
  immunoglobulin rearrangements.
\newblock Bioinformatics. 2010 1~Apr;26(7):867--872.
\newblock Available from:
  \url{http://dx.doi.org/10.1093/bioinformatics/btq056}.

\bibitem{Elhanati2016-yq}
Elhanati Y, Marcou Q, Mora T, Walczak AM.
\newblock {repgenHMM}: a dynamic programming tool to infer the rules of immune
  receptor generation from sequence data.
\newblock Bioinformatics. 2016 26~Feb;Available from:
  \url{http://dx.doi.org/10.1093/bioinformatics/btw112}.

\bibitem{Laserson2012-pi}
Laserson J.
\newblock Bayesian assembly of reads from high throughput sequencing.
\newblock Stanford; 2012.
\newblock Available from: \url{http://purl.stanford.edu/xp796hy4748}.

\bibitem{Laserson2014-yh}
Laserson U, Vigneault F, Gadala-Maria D, Yaari G, Uduman M, Vander~Heiden JA,
  et~al.
\newblock High-resolution antibody dynamics of vaccine-induced immune
  responses.
\newblock Proc Natl Acad Sci U S A. 2014 17~Mar;Available from:
  \url{http://dx.doi.org/10.1073/pnas.1323862111}.

\bibitem{Neal2000-hi}
Neal RM.
\newblock Markov Chain Sampling Methods for {D}irichlet Process Mixture Models.
\newblock J Comput Graph Stat. 2000 1~Jun;9(2):249--265.
\newblock Available from: \url{http://www.jstor.org/stable/1390653}.

\bibitem{Kepler2013-sy}
Kepler TB.
\newblock Reconstructing a {B}-cell clonal lineage. {I}. Statistical inference
  of unobserved ancestors.
\newblock F1000Res. 2013 3~Apr;2:103.
\newblock Available from:
  \url{http://dx.doi.org/10.12688/f1000research.2-103.v1}.

\bibitem{Kepler2014-jy}
Kepler TB, Munshaw S, Wiehe K, Zhang R, Yu JS, Woods CW, et~al.
\newblock Reconstructing a {B}-cell Clonal Lineage. {II}. {M}utation,
  Selection, and Affinity Maturation.
\newblock Front Immunol. 2014 22~Apr;5:170.
\newblock Available from: \url{http://dx.doi.org/10.3389/fimmu.2014.00170}.

\bibitem{Ralph2016-kr}
Ralph DK, Matsen FA IV.
\newblock Consistency of {VDJ} Rearrangement and Substitution Parameters
  Enables Accurate {B} Cell Receptor Sequence Annotation.
\newblock PLoS Comput Biol. 2016 Jan;12(1):e1004409.
\newblock Available from: \url{http://dx.doi.org/10.1371/journal.pcbi.1004409}.

\bibitem{Durbin1998-uq}
Durbin R, Eddy SR, Krogh A, Mitchison G.
\newblock Biological Sequence Analysis: Probabilistic Models of Proteins and
  Nucleic Acids.
\newblock Cambridge University Press; 1998.
\newblock Available from: \url{http://books.google.com/books?id=R5P2GlJvigQC}.

\bibitem{vsearch}
Rognes T.
\newblock Github Repository. 2015;Available from:
  \url{https://github.com/torognes/vsearch}.

\bibitem{Boettiger2014-mm}
Boettiger C.
\newblock An introduction to {D}ocker for reproducible research, with examples
  from the {R} environment. 2014 2~Oct;Available from:
  \url{http://arxiv.org/abs/1410.0846}.

\bibitem{Jiang2011-ur}
Jiang N, Weinstein JA, Penland L, White RA 3rd, Fisher DS, Quake SR.
\newblock Determinism and stochasticity during maturation of the zebrafish
  antibody repertoire.
\newblock Proc Natl Acad Sci U S A. 2011 29~Mar;108(13):5348--5353.
\newblock Available from: \url{http://dx.doi.org/10.1073/pnas.1014277108}.

\bibitem{Vollmers2013-vh}
Vollmers C, Sit RV, Weinstein JA, Dekker CL, Quake SR.
\newblock Genetic measurement of memory {B}-cell recall using antibody
  repertoire sequencing.
\newblock Proc Natl Acad Sci U S A. 2013 13~Aug;110(33):13463--13468.
\newblock Available from: \url{http://dx.doi.org/10.1073/pnas.1312146110}.

\bibitem{Stern2014-ph}
Stern JNH, Yaari G, Vander~Heiden JA, Church G, Donahue WF, Hintzen RQ, et~al.
\newblock {B} cells populating the multiple sclerosis brain mature in the
  draining cervical lymph nodes.
\newblock Sci Transl Med. 2014 6~Aug;6(248):248ra107.
\newblock Available from: \url{http://dx.doi.org/10.1126/scitranslmed.3008879}.

\bibitem{Yaari2015-ss}
Yaari G, Benichou JIC, Vander~Heiden JA, Kleinstein SH, Louzoun Y.
\newblock The mutation patterns in {B}-cell immunoglobulin receptors reflect
  the influence of selection acting at multiple time-scales.
\newblock Philos Trans R Soc Lond B Biol Sci. 2015 5~Sep;370(1676).
\newblock Available from: \url{http://dx.doi.org/10.1098/rstb.2014.0242}.

\bibitem{Gupta2015-sd}
Gupta NT, Vander~Heiden J, Uduman M, Gadala-Maria D, Yaari G, Kleinstein SH.
\newblock {Change-O}: a toolkit for analyzing large-scale {B} cell
  immunoglobulin repertoire sequencing data.
\newblock Bioinformatics. 2015 10~Jun;.

\bibitem{Bolotin2015-zb}
Bolotin DA, Poslavsky S, Mitrophanov I, Shugay M, Mamedov IZ, Putintseva EV,
  et~al.
\newblock {MiXCR}: software for comprehensive adaptive immunity profiling.
\newblock Nat Methods. 2015 29~Apr;12(5):380--381.

\bibitem{billion}
DeWitt WS, Lindau P, Snyder TM, Emerson RO, Sherwood AM, Vignali M, et~al.. A
  public immunosequencing database of memory and na\"ive {B} cell receptors;
  2015.

\bibitem{McCoy2015-qi}
McCoy CO, Bedford T, Minin VN, Bradley P, Robins H, Matsen~IV FA.
\newblock Quantifying evolutionary constraints on {B}-cell affinity maturation.
\newblock Philos Trans R Soc Lond B Biol Sci. 2015 5~Sep;370(1676).
\newblock Available from: \url{http://dx.doi.org/10.1098/rstb.2014.0244}.

\bibitem{Elhanati2015-ld}
Elhanati Y, Sethna Z, Marcou Q, Callan CG Jr, Mora T, Walczak AM.
\newblock Inferring processes underlying {B}-cell repertoire diversity.
\newblock Philos Trans R Soc Lond B Biol Sci. 2015 5~Sep;370(1676).
\newblock Available from: \url{http://dx.doi.org/10.1098/rstb.2014.0243}.

\bibitem{Kuchenbecker2015-hw}
Kuchenbecker L, Nienen M, Hecht J, Neumann AU, Babel N, Reinert K, et~al.
\newblock {IMSEQ} - a fast and error aware approach to immunogenetic sequence
  analysis.
\newblock Bioinformatics. 2015 18~May;.

\bibitem{Bonsignori2011-ky}
Bonsignori M, Hwang KK, Chen X, Tsao CY, Morris L, Gray E, et~al.
\newblock Analysis of a clonal lineage of {HIV-1} envelope {V2/V3}
  conformational epitope-specific broadly neutralizing antibodies and their
  inferred unmutated common ancestors.
\newblock J Virol. 2011 Oct;85(19):9998--10009.
\newblock Available from: \url{http://dx.doi.org/10.1128/JVI.05045-11}.

\bibitem{Liao2013-cr}
Liao HX, Lynch R, Zhou T, Gao F, Alam SM, Boyd SD, et~al.
\newblock Co-evolution of a broadly neutralizing {HIV-1} antibody and founder
  virus.
\newblock Nature. 2013 25~Apr;496(7446):469--476.
\newblock Available from: \url{http://dx.doi.org/10.1038/nature12053}.

\bibitem{Gao2014-ls}
Gao F, Bonsignori M, Liao HX, Kumar A, Xia SM, Lu X, et~al.
\newblock Cooperation of {B} Cell Lineages in Induction of {HIV-1}-Broadly
  Neutralizing Antibodies.
\newblock Cell. 2014 23~Jul;Available from:
  \url{http://dx.doi.org/10.1016/j.cell.2014.06.022}.

\bibitem{Frost2015-rp}
Frost SDW, Murrell B, Hossain ASMM, Silverman GJ, Pond SLK.
\newblock Assigning and visualizing germline genes in antibody repertoires.
\newblock Philos Trans R Soc Lond B Biol Sci. 2015 5~Sep;370(1676).

\bibitem{Volpe2008-xe}
Volpe JM, Kepler TB.
\newblock Large-scale analysis of human heavy chain {V(D)J} recombination
  patterns.
\newblock Immunome Res. 2008 27~Feb;4:3.
\newblock Available from: \url{http://dx.doi.org/10.1186/1745-7580-4-3}.

\bibitem{Kidd2015-vt}
Kidd MJ, Jackson KJL, Boyd SD, Collins AM.
\newblock {DJ} Pairing during {VDJ} Recombination Shows Positional Biases That
  Vary among Individuals with Differing {IGHD} Locus Immunogenotypes.
\newblock J Immunol. 2015 23~Dec;Available from:
  \url{http://dx.doi.org/10.4049/jimmunol.1501401}.

\bibitem{Saada2007-kf}
Saada R, Weinberger M, Shahaf G, Mehr R.
\newblock Models for antigen receptor gene rearrangement: {CDR3} length.
\newblock Immunol Cell Biol. 2007 3~Apr;85(4):323--332.
\newblock Available from: \url{http://dx.doi.org/10.1038/sj.icb.7100055}.

\bibitem{Kepler1996-kd}
Kepler TB, Borrero M, Rugerio B, McCray SK, Clarke SH.
\newblock Interdependence of {N} nucleotide addition and recombination site
  choice in {V(D)J} rearrangement.
\newblock The Journal of Immunology. 1996 15~Nov;157(10):4451--4457.
\newblock Available from:
  \url{http://www.jimmunol.org/content/157/10/4451.abstract}.

\bibitem{Jackson2007-ue}
Jackson KJL, Ga{\"{e}}ta BA, Collins AM.
\newblock Identifying highly mutated {IGHD} genes in the junctions of
  rearranged human immunoglobulin heavy chain genes.
\newblock J Immunol Methods. 2007 15~May;324(1-2):26--37.
\newblock Available from: \url{http://dx.doi.org/10.1016/j.jim.2007.04.011}.

\bibitem{Larimore2012-lo}
Larimore K, McCormick MW, Robins HS, Greenberg PD.
\newblock Shaping of human germline {IgH} repertoires revealed by deep
  sequencing.
\newblock J Immunol. 2012 3~Aug;189(6):3221--3230.
\newblock Available from: \url{http://dx.doi.org/10.4049/jimmunol.1201303}.

\bibitem{MAL-013}
Sutton C, McCallum A.
\newblock An Introduction to Conditional Random Fields.
\newblock Foundations and Trends in Machine Learning. 2011;4(4):267--373.
\newblock Available from: \url{http://dx.doi.org/10.1561/2200000013}.

\bibitem{Gadala-Maria2015-uq}
Gadala-Maria D, Yaari G, Uduman M, Kleinstein SH.
\newblock Automated analysis of high-throughput {B}-cell sequencing data
  reveals a high frequency of novel immunoglobulin {V} gene segment alleles.
\newblock Proc Natl Acad Sci U S A. 2015 9~Feb;.

\bibitem{Siepel2005-mw}
Siepel A, Haussler D.
\newblock Phylogenetic Hidden Markov Models.
\newblock In: Statistical Methods in Molecular Evolution. Statistics for
  Biology and Health. Springer New York; 2005. p. 325--351.
\newblock Available from:
  \url{http://link.springer.com/chapter/10.1007/0-387-27733-1_12}.

\bibitem{Wilson2000-bg}
Wilson PC, Wilson K, Liu YJ, Banchereau J, Pascual V, Capra JD.
\newblock Receptor revision of immunoglobulin heavy chain variable region genes
  in normal human {B} lymphocytes.
\newblock J Exp Med. 2000 5~Jun;191(11):1881--1894.
\newblock Available from: \url{http://www.ncbi.nlm.nih.gov/pubmed/10839804}.

\bibitem{DeKosky2013-iz}
DeKosky BJ, Ippolito GC, Deschner RP, Lavinder JJ, Wine Y, Rawlings BM, et~al.
\newblock High-throughput sequencing of the paired human immunoglobulin heavy
  and light chain repertoire.
\newblock Nat Biotechnol. 2013 Feb;31(2):166--169.
\newblock Available from: \url{http://dx.doi.org/10.1038/nbt.2492}.

\bibitem{McDaniel2016-db}
McDaniel JR, DeKosky BJ, Tanno H, Ellington AD, Georgiou G.
\newblock Ultra-high-throughput sequencing of the immune receptor repertoire
  from millions of lymphocytes.
\newblock Nat Protoc. 2016 Mar;11(3):429--442.
\newblock Available from: \url{http://dx.doi.org/10.1038/nprot.2016.024}.

\bibitem{Price2010-fi}
Price MN, Dehal PS, Arkin AP.
\newblock {FastTree} 2--approximately maximum-likelihood trees for large
  alignments.
\newblock PLoS One. 2010;5(3):e9490.
\newblock Available from:
  \url{http://dx.plos.org/10.1371/journal.pone.0009490.g003}.

\bibitem{Goldman2000-qw}
Goldman N, Anderson JP, Rodrigo AG.
\newblock {Likelihood-Based} Tests of Topologies in Phylogenetics.
\newblock Syst Biol. 2000 Dec;49(4):652--670.
\newblock Available from: \url{http://dx.doi.org/10.1080/106351500750049752}.

\bibitem{Stadler2011-sz}
Stadler T.
\newblock Simulating trees with a fixed number of extant species.
\newblock Syst Biol. 2011 Oct;60(5):676--684.
\newblock Available from: \url{http://dx.doi.org/10.1093/sysbio/syr029}.

\bibitem{bppseqgen}
J D, B B.
\newblock Non-homogeneous models of sequence evolution in the {Bio++} suite of
  libraries and programs.
\newblock BMC Evol Biol. 2008 Sep;8(1):255.

\end{thebibliography}

\clearpage
\beginsupplement

\section*{Supplementary Information}

\end{document}